\documentclass[11pt]{mart_jksu}
\usepackage{cite}
\usepackage{epsf}
\usepackage{times}
\usepackage{mathptm}

\begin{document}

\vspace{1cm}

\begin{center}
{\LARGE From Pauli Principle to Hypernucleus, Neutron Star, 

\vspace{0.2cm}

and Econophysics}

\vspace{1cm}

{\bf T. Mart}


{\sl Departemen Fisika, FMIPA, Universitas Indonesia, Depok 16424, 
  Indonesia
}

\vspace{0.4cm}

({\today})

\end{center}

\vspace{0.4cm}

\begin{abstract}
Proposed by Wolfgang Pauli more than 80 years ago, the exclusion 
principle has been proven to have a far-reaching consequence, 
from femtoscopic world to macroscopic, super-dense,
and fully relativistic physics. Starting from this principle, 
we discuss two interesting research topics, which have currently 
drawn considerable attention in the nuclear- and astrophysics communities;
the hypernuclear and neutron star physics. Special attention is
given to the electromagnetic production of the hypertriton and 
the consequences of the neutrino electromagnetic properties in
dense matter. We also touch on the new arena 
which could also be fascinating for physicists; the econophysics.
\end{abstract}


\section{Pauli Exclusion Principle and Its Consequence}
\subsection{Historical Background}

Most of us started to be acquainted with the Pauli exclusion
principle at high-school, when we learned about atomic shells in the 
chemistry or in the physics course. Probably, many of us 
are ignorant that the discovery of the exclusion principle is 
in fact intimately related to the conception of 
the electron spin and the development of quantum mechanics. If we try to comprehend 
the motivation of Wolfgang Pauli behind this great discovery, for
which he was awarded with the Nobel Prize in 1945, then we will
find that it was to answer the problem of explaining the observed 
atomic spectra and the ``anomalous Zeeman effect'' 
\cite{pauli_nobel}.
At that time these transitions were particularly confusing,
because some of them (e.g., the Sodium atom) exhibit doublet 
transitions, whereas some others (e.g., the Magnesium atom) show
either singlet or triplet transitions  \cite{martin}. 
Furthermore, the anomalous Zeeman effect yields a doublet splitting, contrary
to the what was hitherto known by the spectroscopists as triplet splitting. 
Another disturbing fact to Pauli was the question as to why all electrons
for an atom in its ground state were not bound in the innermost
shell, a problem which had been emphasized even by Niels Bohr as 
his fundamental problem. 

Before 1924, Arnold Sommerfeld, Alfred Land\'e, and Pauli himself realized that
the answer to these problems was the existence of a new (the fourth,
after $n, \ell$ and $m$,
to be more precise) quantum number. Sommerfeld and Land\'e considered
this number as an intrinsic property of the atomic
core, without specifying whether it is the nucleus or the nucleus
plus inner electrons. This point of view was then considered by
Pauli as rather orthodox.

In 1924 Pauli published an objection to this argument \cite{pauli1924}.
Instead of considering the properties of the atomic core, 
Pauli proposed a new 
quantum property of the electron, which he called 
``two-valuedness not describable classically.'' Using this idea
Pauli was able to reduce the complicated numbers of electrons in closed
subgroups to the simple one, once this fourth quantum property is 
introduced. Nevertheless, with the exception of few famous 
spectroscopists, nobody could easily follow this idea, since there was
no analogy in terms of mechanical model. This difficulty was then
remedied by the concept of electron spin proposed by George E. Uhlenbeck and
Samuel A. Goudsmit, who tried to explain the experimental result of
Otto Stern and Walther Gerlach \cite{stern_gerlach}. 
It was said that after some discussions with Lorentz, they 
tried to withdraw their paper, but it was too
late and the paper was published! With this concept, i.e. by assuming 
that the spin quantum number of electron as $\frac{1}{2}$, it is then 
possible to understand the doublet splitting in the anomalous
Zeeman effect \cite{uhlenbeck}.
Since then the idea of exclusion principle 
was always closely connected with the spin. 

\begin{figure}[t]
\centerline{\epsfxsize=3cm \epsffile{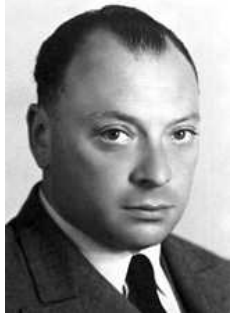}}
\caption{\small Wolfgang Pauli (1900-1958), the 1945 Nobel laureate 
  for ``the discovery of the exclusion principle''
  \cite{noble_homepage}.}
   \label{fig:pauli} 
\end{figure}

Pauli formulated his exclusion principle comprehensively in 1925
\cite{pauli1925}. At about the same time, several milestones in 
quantum mechanics were achieved. The most important and 
relevant one to the present discussion is the formulation
of the quantum mechanics by Werner Heisenberg \cite{heisenberg}.
According to Heisenberg, quantum mechanics leads to qualitatively different 
conclusions for distinguishable particles than for indistinguishable
particles. It was found that for indistinguishable
particles the wave function must be either
symmetric or antisymmetric. Pauli considered the antisymmetric one
as the appropriate wave mechanical formulation of the 
exclusion principle, since an antisymmetric wave function containing 
two particles in the same state is equal to zero. 

The first indication that proton has spin $\frac{1}{2}$ was obtained from
the investigation of the anomaly in the specific heat of the 
molecular hydrogen in 1927 \cite{dennison}. 
Later on, this was confirmed by Stern by using a Stern-Gerlach
type experiment. In 1932, the neutron was discovered by Frederic and
Ir\`ene Joliot-Curie along with James Chadwick. Two years later, 
by dissociating deuteron with gamma ray, $\gamma + d \to n + p$, 
Chadwick and Maurice Goldhaber were 
able to show that a deuteron was made of a proton and a neutron.
From its spectrum it was known that the deuteron has spin 1. 
Hence, we know that the neutron has also 
spin  $\frac{1}{2}$ and both proton and neutron obey the Pauli exclusion principle.

Although for us the exclusion principle is quite clear and understandable,
Pauli himself was disappointed with his finding, since he said that
he was unable to give a logical reason or to deduce the principle
from a more general assumption. He was hoping that the development
of quantum mechanics will also rigorously deduce the principle, but
he eventually found that it was not the case \cite{pauli_nobel}.

\subsection{The Degeneracy Force}

\begin{figure}[t]
\centerline{\epsfxsize=7cm \epsffile{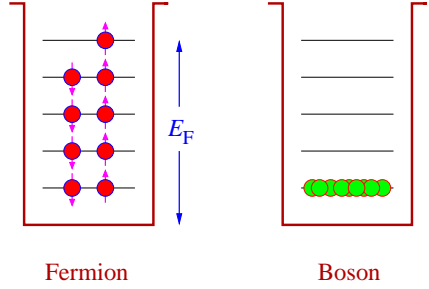}}
\caption{\small If nine fermions and nine bosons are ``dropped''
  into square well potentials, they will organize themselves
  in quite different ways. This happens because the fermions 
  always obey the Pauli exclusion principle.}
   \label{fig:fermion_boson} 
\end{figure}

Let us begin with a simple gedanken experiment, i.e., we drop
a number of fermions into a square well potential. We repeat
the same experiment for a number of bosons. As we expected, 
the result is shown in Fig.~\ref{fig:fermion_boson}. However, 
to understand why the fermions sitting above the ground state
do not collapse to the lowest state as in the case of bosons,
we need to understand the degeneracy force, a kind of
force which maintains such formation. 

Consider two particles $a$ and $b$ located at 
$x_1$ and $x_2$. The wave function of such a 
system can be written as \cite{griffiths}
\begin{eqnarray}
  \label{eq:tidak-berkorelasi}
  \Psi (x_1,x_2) &=& \psi_a(x_1)\, \psi_b(x_2) ~,
\end{eqnarray}
where both $\psi_a$ and $\psi_b$ are normalized to unity.
Let us now calculate the expectation value
of the ``distance'' between the two particles, i.e.,
\begin{eqnarray}
  \label{eq:distance}
  \langle (x_1-x_2)^2 \rangle ~=~ \langle x_1^2 \rangle +   
  \langle x_2^2 \rangle -2  \langle x_1x_2 \rangle ~.
\end{eqnarray}
Using the wave function given in Eq.\,(\ref{eq:tidak-berkorelasi})
we can calculate the first term on the r.h.s. of 
Eq.\,(\ref{eq:distance}),
\begin{eqnarray}
  \langle x_1^2 \rangle ~=~ \int dx_1\, x_1^2\, |\psi_a(x_1)|^2\,
  \underbrace{\int dx_2\,|\psi_b(x_2)|^2}_{=1}
  ~=~ \int dx~ x^2\, |\psi_a(x)|^2\, ~\equiv~
  \langle x^2 \rangle_a ~,
\end{eqnarray}
where the subscript $a$ 
indicates that the expectation
value is calculated using the wave function 
$\psi_a(x)$. 
By following the same procedure we can easily prove that 
$\langle x_2^2 \rangle = \langle x^2 \rangle_b$ and
$\langle x_1x_2 \rangle = \langle x \rangle_a\langle x \rangle_b$.
Therefore, the ``distance `` between two distinguishable particles is
found to be 
\begin{eqnarray}
  \label{eq:distance-distinguish}
  \langle (x_1-x_2)^2 \rangle_{\rm dis} ~=~ 
    \langle x^2 \rangle_a +   \langle x^2 \rangle_b -2  
    \langle x\rangle_a \langle x \rangle_b ~.
\end{eqnarray}

 On the other hand, the wave function for two indistinguishable boson or
fermion is given by
\begin{eqnarray}
  \label{eq:indistinguishable}
  \Psi_{\pm}(x_1,x_2) ~=~ \frac{1}{\sqrt{2}}\left[\psi_a(x_1)\, \psi_b(x_2)\pm\psi_a(x_2)\, \psi_b(x_1)\right] ~,
\end{eqnarray}
where the $+$ and $-$ signs indicate the symmetric (boson) and anti-symmetric
(fermion) systems.
Using Eq.\,(\ref{eq:indistinguishable}) we can recalculate 
Eq.\,(\ref{eq:distance-distinguish}) and obtain
\begin{eqnarray}
  \label{eq:distance-indistinguish}
  \langle (x_1-x_2)^2 \rangle_\pm ~=~ 
    \langle x^2 \rangle_a +   \langle x^2 \rangle_b -
    2 \langle x \rangle_a\langle x \rangle_b
    \,\mp\, 2\, |\langle x \rangle_{ab}|^2~,
\end{eqnarray}
where the last term, 
$\langle x \rangle_{ab}=\int dx\,  \psi_a^*(x)\, x\, \psi_b(x)$, 
clearly depends upon the overlap between 
the wave functions $\psi_a(x)$ and $\psi_b(x)$.

 By comparing Eqs. (\ref{eq:distance-distinguish}) and
(\ref{eq:distance-indistinguish}) we can conclude that
\begin{eqnarray}
  \label{eq:distance-indistinguish-distinguish}
  \langle (x_1-x_2)^2 \rangle_\pm ~=~ \langle (x_1-x_2)^2 \rangle_{\rm dis}
    \,\mp\, 2\, |\langle x \rangle_{ab}|^2~.
\end{eqnarray}
This equation is extremely important. Although derived in a quite
simple fashion, it shows that two fermions tend to be farther apart 
than two bosons. This is known as the exchange force (or degeneracy
force), which produces a ``force of attraction'' between identical
bosons and a ``force of repulsion'' between identical fermions. 
Note that this is not really the force we are familiar with,
since there is no mediator which pull or push the particles,
instead it is solely a consequence of the (anti)symmetrization
required by quantum mechanics. 

After we know that identical fermions exhibit ``repulsive force'', our
next question is certainly ``how repulsive is this force?'' To answer
this question we use the electron gas model, since in this model
electrons do not interact via Coulomb force. Furthermore, the 
electron gas model is a simple model often used in many standard
quantum mechanics books \cite{gasiorowicz}. 
In this model, 
for $N$ number of electrons, each with a mass of $m$, occupying 
a volume $V$, the corresponding total energy can be written as 
\begin{eqnarray}
    E_{\rm tot.} ~=~ \frac{\hbar^2\pi^3}{10mV^{2/3}}\left( \frac{3N}{\pi}
    \right)^{5/3} ~.
\end{eqnarray}
Using the density $\rho=N/V$ we can calculate the ``pressure'' produced
by the degeneracy force of $N$ electrons and find that 
\begin{eqnarray}
    \label{eq:degenerate_force1}
    p_{\rm deg.} ~=~ -\frac{\partial E_{\rm tot.}}{\partial V}
    ~=~ \frac{\hbar^2\pi^3}{15m}\left( \frac{3\rho}{\pi}
    \right)^{5/3}~.
\end{eqnarray}
To get a better feeling on this pressure, let us consider two ``chargeless''
electrons separated with a distance of 1 \AA. The particle density in
this case is $\rho\approx 10^{30}$ m$^{-3}$. Substituting this value
into Eq.\,(\ref{eq:degenerate_force1}) we obtain the degeneracy
pressure is $p_{\rm deg.}\approx 10^{14}$ N/m$^2$. One may suspect
that such a huge pressure would be meaningless in an atomic scale,
where the typical ``area'' is of the order of $10^{-20}$ m$^2$ and therefore
the corresponding force is of the order of $10^{-6}$ N. Nevertheless, one
should not forget that at this scale the Coulomb force between two electrons is also of the
order of $10^{-8}$ N. The problem, however, turns out to be dramatically interesting
on the macroscopic scale where we do business with an 
area greater than 1 km$^2$ or $10^6$~m$^2$. We will encounter this
interesting problem when we discuss the neutron star, but first we 
will introduce the physics of hypernucleus.

\begin{figure}[t]
\centerline{\epsfxsize=12cm \epsffile{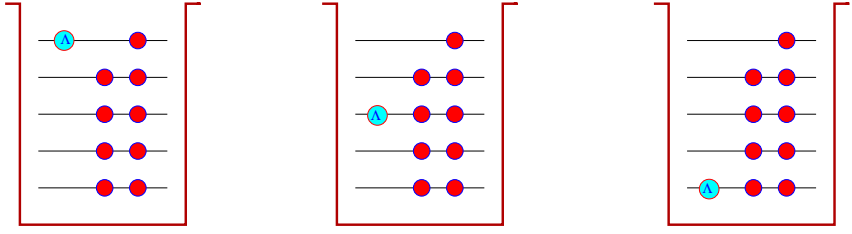}}
\caption{\small A $\Lambda$-hyperon can freely move in the nuclear
  energy states, without being blocked by the Pauli principle.}
   \label{fig:lambda} 
\end{figure}

\section{Introducing a $\Lambda$ Hyperon in the Nucleus; Hypernuclear Physics}
\subsection{Pauli Principle in the Nucleus}
Let us now imagine that we can implant a $\Lambda$ hyperon inside a normal
nucleus. Since the $\Lambda$ hyperon is distinguishable from the nucleons
(the protons and neutrons which build up the nucleus), it will certainly be 
free from Pauli blocking (see Fig.~\ref{fig:lambda}). Such a nucleus is
called a hypernucleus. Hypernucleus is interesting because many new 
information about the nuclear structure can be obtained by 
studying the way the $\Lambda$ hyperon localizes inside the nucleus.
Another interesting aspect comes from the fact that a free $\Lambda$ hyperon
is unstable, it decays in about $10^{-10}$ s primarily to 
\begin{eqnarray}
  \Lambda ~\to~ \left\{
  \begin{array}[l]{lr}
    p ~+~ \pi^- & ({\rm branching~fraction~:}~63.9\%)\\
    n ~+~ \pi^0 & ({\rm branching~fraction~:}~35.8\%)
  \end{array} \right. ~,
\label{eq:lambda_decay}
\end{eqnarray}
via weak interaction. However, if the $\Lambda$ is embedded in a nucleus, 
since the low energy nucleon states in the nucleus are already filled,
Pauli principle will prevent it from decaying in the usual way as in 
Eq.~(\ref{eq:lambda_decay}). Therefore, there must be another mechanism
which eventually allows the hypernucleus to decay via a weak process.

\begin{figure}[t]
\centerline{\epsfxsize=5cm \epsffile{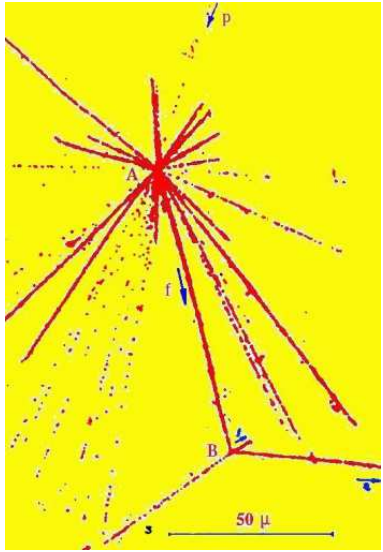}}
\caption{\small The first observed hypernucleus. 
  A highly energetic proton $p$ strikes a nucleus at point A in an emulsion,
  which immediately breaks up into a number of nuclear fragments. All fragments
  are eventually stopped in the emulsion except the one marked with f. This fragment 
  decays at point B into three charged particles and a neutron. The neutron is not
  seen in this figure, since it is chargeless.
  Original figure was taken from Ref.~\cite{Danysz_Pniewski}.}
   \label{fig:hypernuleus} 
\end{figure}

 The first hypernucleus was observed by Marian Danysz and Jerzy Pniewski, 
two Polish physicists, in 1953. They used photographic emulsions exposed
to the cosmic radiation. These emulsions recorded a break-up event illustrated
in Fig.~\ref{fig:hypernuleus}. We can understand this event as follows.
A highly energetic proton coming from cosmic rays hits a heavy nucleus
in the emulsion, causing it to break up into a number of fragments. One of them
(marked with f) decays into three charged particles at point B. This decay 
cannot be interpreted as a normal decay of a highly excited nucleus, since
from the track shown in the figure the three particles should have lived for
about $10^{-12}$ s, whereas the normal decay of an excited nucleus would take 
place in about
$10^{-22}$ s. Presumably, the event can be interpreted as the decay of the
$^8_\Lambda$Be hypernucleus into three charged particles \cite{hodgson}, 
\begin{eqnarray}
  ^8_\Lambda{\rm Be} ~\to~ ^3{\rm He} ~+~ ^4{\rm He} ~+~ n ~+~ \pi^- ~.
\end{eqnarray}

 The first hypernucleus was studied in a photographic emulsion exposed to 
cosmic rays. Nowadays, hypernuclei can be produced in laboratories by 
using, e.g., hadronic reactions
\begin{eqnarray}
  \label{eq:hadronic_hyp}
  K^- ~+~ ^AZ ~\to~ ^A_\Lambda{Z} ~+~ \pi^- ~~~,~~~~~
  \pi^+ ~+~ ^AZ ~\to~ ^A_\Lambda{Z} ~+~ K^+ ~,
\end{eqnarray}
or through electromagnetic interactions
\begin{eqnarray}
  \label{eq:electromagnetic_hyp}
  \gamma ~+~ ^AZ ~\to~ ^A_\Lambda{Z} ~+~ K^+ ~~~,~~~~~
  e ~+~ ^AZ ~\to~ e' ~+~ ^A_\Lambda{Z} ~+~ K^+ ~.
\end{eqnarray}
The advantage of using the first reactions, Eq.~(\ref{eq:hadronic_hyp}), 
is that the cross section is large and therefore the corresponding 
experiments are relatively easy to perform. However,
the electromagnetic reactions, Eq.~(\ref{eq:electromagnetic_hyp}), are also
invaluable for the study of $\Lambda$ hypernuclei. Since a large momentum 
is transferred to the recoiled hypernucleus, it will then populate high
spin states and, therefore, is very suitable for investigating deeply
bound states of a $\Lambda$ hyperon. Furthermore, both spin-flip and non-spin-flip
amplitudes are excited with significant cross sections, in contrast to
the hadronic reactions. In addition, since a proton is converted to a $\Lambda$
hyperon, hypernuclei that are not accessible by hadronic reactions can
be produced. Finally, a comparison between hadronic and electromagnetic
reactions allows us to study charge symmetry in the $\Lambda$ hypernuclei \cite{hashimoto}.

\begin{figure}[t]
\centerline{\epsfxsize=7cm \epsffile{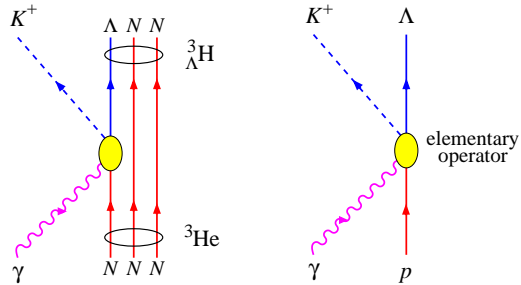}}
\caption{\small {\bf (Left)} Photoproduction of a hypertriton. {\bf (Right)}
  Photoproduction of a $\Lambda$ hyperon which is called elementary reaction.
  The amplitude obtained from Feynman diagrams for this reaction is called
  elementary operator.}
   \label{fig:hypertriton} 
\end{figure}

 The simplest example is photoproduction of a hypertriton, the lightest hypernucleus.
A hypertriton consists of two nucleons and one $\Lambda$ hyperon. Thus, it can be
produced by bombarding a $^3$He target with (real or virtual) photon, as 
shown in Fig.~\ref{fig:hypertriton}. From this figure it is clear that one of
the proton inside the nucleus is converted into a $\Lambda$ hyperon along with a kaon
through an elementary process. Therefore, to theoretically calculate 
the hypertriton photoproduction one needs information about the elementary reaction. 

\subsection{Elementary Reaction}
\begin{figure}[t]
\centerline{\epsfxsize=12cm \epsffile{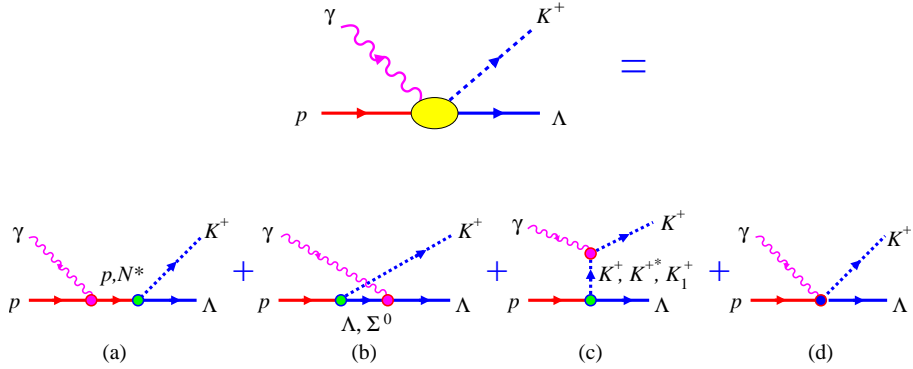}}
\caption{\small Elementary operator of $K\Lambda$ photoproduction on the nucleon
  can be modeled by several Feynman diagrams at tree-level.
  Diagrams (a), (b), and (c) correspond to the $s$-, $u$-, and $t$-channel,
  respectively, whereas diagram (d) represents the contact term which is
  required to maintain gauge invariance of the process.}
   \label{fig:elementary} 
\end{figure}

The elementary operator can be modeled in terms of Feynman 
diagrams at tree-level as shown in Fig.~\ref{fig:elementary}. 
Although higher-order corrections are neglected, most of the hadronic 
and electromagnetic couplings in these diagrams are in fact not well known. 
Fortunately, experimental data for the elementary process 
$\gamma + p\to K^+ + \Lambda$ have been available since more than
40 years ago. Thus, all unknown coupling constants can be extracted by
fitting the calculated observables to these data. 

\begin{figure}[h]
\begin{minipage}[h]{7cm}
{\epsfxsize=7cm \epsffile{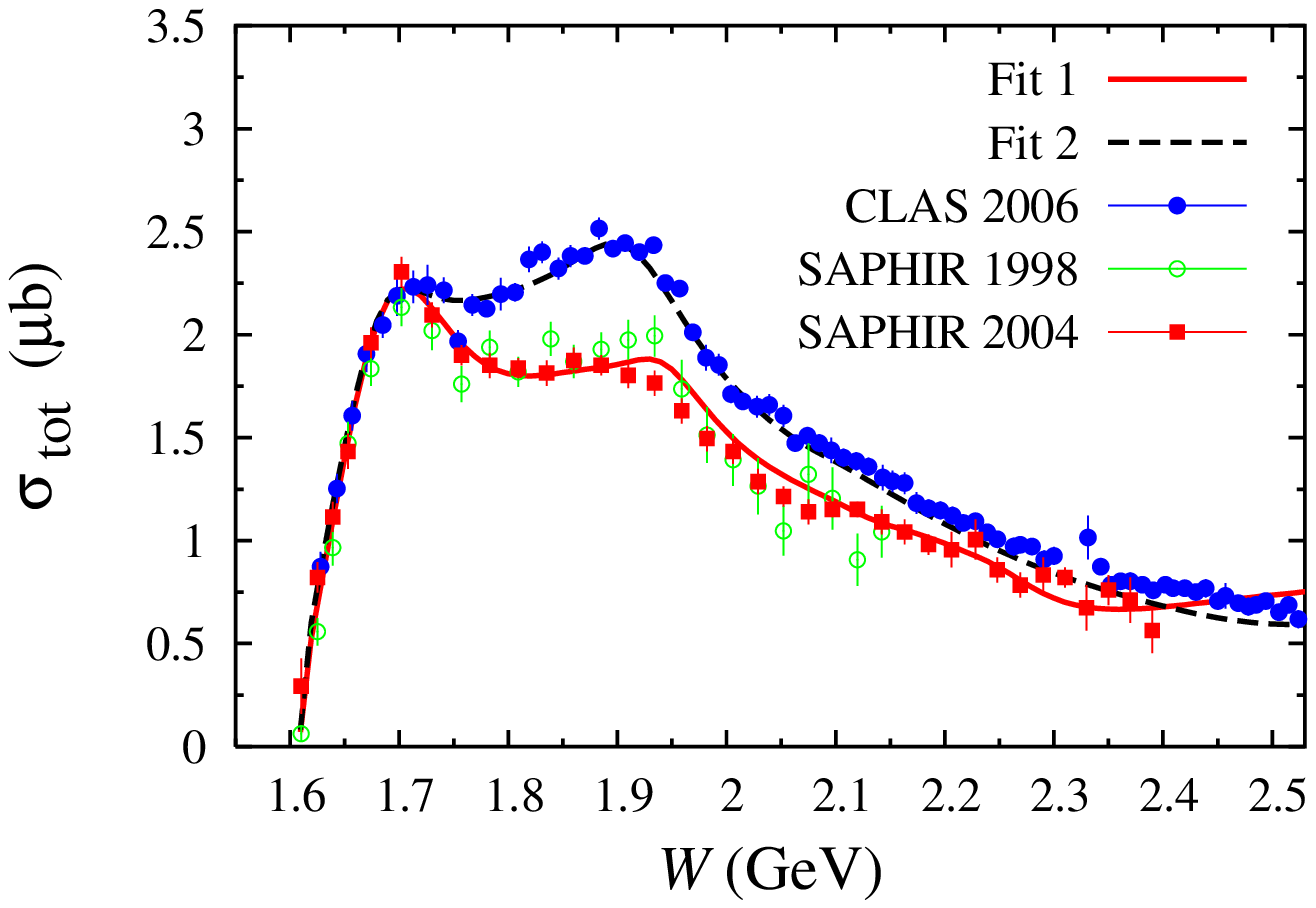}}
\end{minipage}
\hspace{\fill}
\begin{minipage}[h]{9cm}
{\epsfxsize=9cm \epsffile{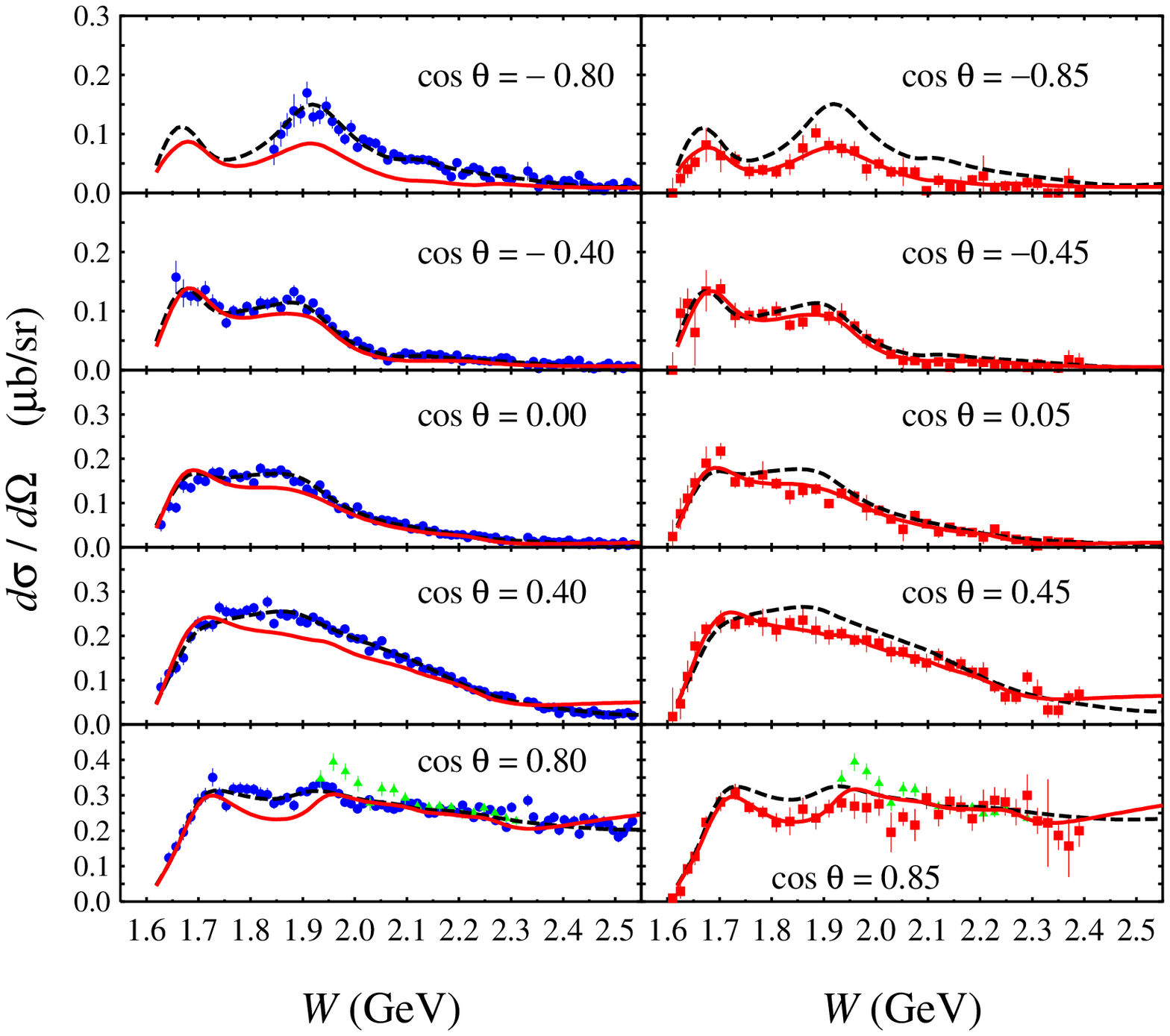}}
\end{minipage}
\caption{\small {\bf (Left)} Comparison between experimental total cross sections 
  with the predictions of Fit 1 (fit to the SAPHIR data) and Fit 2 (fit to the CLAS data).
    {\bf (Right)} Differential cross sections obtained from Fit 1
  and Fit 2 as a function of the total c.m. energy for different kaon angles.
  Solid circles, squares, and triangles represent experimental data
  from the {CLAS}, {SAPHIR}, and {LEPS} collaborations, respectively \cite{Mart:2006dk}.}
  \label{fig:elementary_dcs}
\end{figure}

 Very recently a large number of data have been published by the SAPHIR
\cite{Glander:2003jw} and CLAS \cite{Bradford:2005pt} collaborations. 
These data cover the energy range from reaction threshold up to
$W=2.5$ GeV, angular distributions from forward to backward kaon
angles, and are significantly more accurate than the previous ones
\cite{Tran:1998qw}. A comparison between theoretical calculations 
with small part of these data is shown in Fig.~\ref{fig:elementary_dcs}.
To calculate the differential cross section we have used 
\begin{eqnarray}
  \label{eq:cs_nucleon}
\frac{d\sigma}{d\Omega} 
&=& \frac{|\vec{q}|}{|{\vec k}|}~
\frac{m_p E_{\Lambda}}{32\pi^2W^2}~\sum_{\epsilon}
|{\cal M}_{\rm fi}|^2 ~,
\end{eqnarray}
where $\vec{q}$ and ${\vec k}$ are the kaon and photon momenta, respectively,
whereas $W$ is the total c.m. energy. The matrix element ${\cal M}_{\rm fi}$ is
obtained from the Feynman diagrams shown in Fig.~\ref{fig:elementary} and
is summed over the possible photon polarization $\epsilon$.

 A quick glance to Fig.~\ref{fig:elementary_dcs} can immediately convince
us that the new data have a problem of the mutual consistency. To study
the physics consequence of this problem, Refs.~\cite{Bydzovsky:2006wy}
and \cite{Mart:2006dk} have performed different fits to several data 
sets. In Fig.~\ref{fig:elementary_dcs} we show only two of them, i.e.,
fit to the new SAPHIR data (Fit 1) and fit to the new CLAS data (Fit 2).
The new LEPS data are found to be more consistent with the CLAS ones.

 Reference~\cite{Mart:2006dk} has pointed out that the use of SAPHIR and 
CLAS data, individually or simultaneously, leads to quite different 
resonance parameters. This could lead to different conclusions 
on ``missing resonances'', the resonances which have been predicted by 
constituent quark models but are intangible to $\pi +N \to \pi +N$ 
reactions that are used by the Particle Data Group (PDG) to extract 
their properties. 

In view of this, the elementary process of kaon production can also
be used as a complimentary tool for
investigation of the ``missing resonances''. As an example, it 
is worth mentioning the case of ``missing'' $D_{13}(1895)$ which was observed 
by Ref.~\cite{Mart:1999ed} in the $\gamma + p \to K^+ +\Lambda$ 
channel. It was found after the 
previous SAPHIR data \cite{Tran:1998qw} showed a clear peak
around $W=1900$ MeV in the differential and total cross sections.
Subsequently, it was shown in Ref.~\cite{Janssen:2001wk} that
the peak could also be equally well reproduced by including a $P_{13}(1950)$
resonance. However, most of analyses based on the isobar model after that
confirmed that including the $D_{13}(1895)$ significantly improves the
agreement between theoretical prediction and 
experimental data. Recent partial-wave and coupled-channels 
analyses confirmed that the peak originates from a $D_{13}$ resonance
with a mass around 1900 MeV \cite{many_studied_D13}. 
Indeed, the same conclusion has also been drawn from the recent multipole 
study~\cite{Mart:1999ed}, that the peak at $W\sim 1900$ MeV 
originates from a $D_{13}$  resonance with the extracted mass 
equals 1936 MeV if SAPHIR data were used or 1915 MeV if CLAS data were used.
Thus, future measurements such as the one planned at MAMI in Mainz are, 
therefore, expected to remedy this unfortunate situation.

\subsection{Electromagnetic Production of the Hypertriton}
Theoretical calculation of photoproduction of the hypertriton can be done 
by making use of the left diagram of Fig.~\ref{fig:hypertriton}. The formula
for calculating the hypernuclear cross section 
is similar to the one for calculating the elementary cross
sections [cf. Eq.~(\ref{eq:cs_nucleon})], i.e.,
\begin{eqnarray}
  \label{eq:cs_hypertriton}
  \frac{d\sigma}{d\Omega} &=& \frac{|\vec{q}|}{|\vec{k}|}
~\frac{M_{\rm ^3He} E_{\rm ^3_{\Lambda}H}}{32\pi^2W^2}
~\sum_{\epsilon} \sum_{M,M'}~{\textstyle \frac{1}{2}}\left|{\cal T}_{\rm fi}\right|^2 ~,
\end{eqnarray}
except now the nuclear amplitude ${\cal T}_{\rm fi}$ squared is averaged over
the initial and final nuclear spins. The nuclear amplitude is somewhat complicated
in this case. It consists of the elementary part, integrated over the possible initial
and final nucleon momenta and weighted by the nuclear wave functions, and 
some angular parts in terms of $6j$ and $9j$ coefficients. Explicitly, it is
given by
\begin{eqnarray}
T_{\mathrm fi} & = &  \sqrt{\frac{6}{\pi}} 
~\sum_{\alpha {\alpha}'} ~\sum_{n \Lambda  m_{\Lambda}}
i^{n} \hat n \hat {{\cal L}} \hat {{\cal S}} \hat {{\cal S}'} \hat \Lambda 
(-)^{1 + n + {{\cal S}} + M} \delta_{SS'} \delta_{LL'} \delta_{T0} \times 
\nonumber\\ 
&& \left( \begin{array}{ccc} \frac{1}{2} & \frac{1}{2} & \Lambda\\
M' & -M & m_{\Lambda} \end{array} \right)
\left\{ \begin{array}{ccc} {{\cal S}} & {{\cal S}'} & n \\ \frac{1}{2} & 
\frac{1}{2} & 1 \end{array} \right\}
\left\{ \begin{array}{ccc} {{\cal L}} & {{\cal S}} & \frac{1}{2} \\ 
L & {{\cal S}'} & \frac{1}{2} \\
l & n & \Lambda \end{array} \right\} \times \nonumber\\
&& \int d^{3}{\vec q}~ p^{2}dp~ 
\varphi_{\Lambda}(q\, ')~ \Psi_{d}^{(L)}(p)~ \phi_{\alpha}(p,q)~ 
\left[ {\bf Y}^{(l)} (\hat{{\vec q}\, }) \otimes {\bf K}^{(n)} 
\right]^{(\Lambda )}_{m_{\Lambda}} ~, 
\label{eq:tfi}
\end{eqnarray}
where the amplitude ${\bf K}^{(n)}$ is related to the elementary
amplitude given in Eq.~(\ref{eq:cs_nucleon}) by
\begin{eqnarray}
{\cal M}_{\rm fi} &=& L + i\, {\vec{\sigma}} \cdot {\vec K} ~=~ 
 \sum_{n=0,1} (-i)^{n~} \hat n ~\left[ {\bf \sigma}^{(n)} \otimes 
{\bf K}^{(n)} \right]^{(0)} ~,
\label{eq:kln_def}
\end{eqnarray}
with $\sigma^{(n)}$ is the Pauli matrices. 

\begin{figure}[htb]
\centerline{\epsfxsize=6cm \epsffile{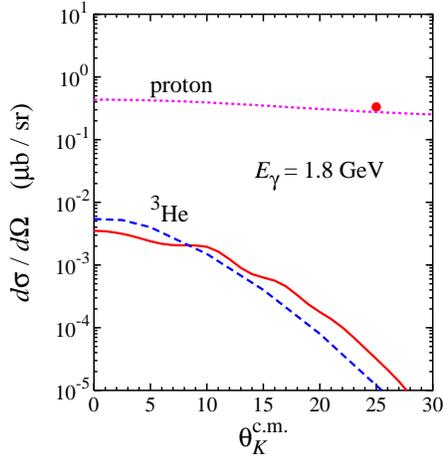}}
\caption{\small Differential cross section for kaon photoproduction off the proton
  (dotted line) and $^3$He (dashed and solid lines) as a function of kaon angle.  
  The dashed line shows the ``factorization'' approximation for production off the $^3$He, whereas 
  the solid line represents the exact calculation by using $S$-waves
  \cite{Mart:1996ay}. Solid circle represents the CLAS datum at this kinematic
  \cite{Bradford:2005pt}.}
   \label{fig:ts_fq2} 
\end{figure}

 The result is shown in Fig.~\ref{fig:ts_fq2}, where we compare the predicted cross sections 
of hypertriton production with that of elementary reaction. For the hypernuclear
production we have performed an exact calculation using Eq.~(\ref{eq:cs_hypertriton})
and an approximated one by using the so-called ``factorization'' approximation.
In the latter, the cross section is calculated by means of
\begin{eqnarray}
  \label{eq:approximated_cs}
 \frac{d\sigma}{d\Omega} &=&{\textstyle \frac{1}{6}}~ W_A^2~|F(Q)|^2 
~\left( \frac{d\sigma}{d\Omega}\right)_{\rm elementary} ~,
\end{eqnarray}
where $W_A$ is a kinematical factor and the ``form factor'' $F(Q)$ is given by
the overlap integral of the two wave functions, 
\begin{eqnarray}
  \label{eq:efq1}
  F(Q) &=& \left. \left\langle ^3_{\Lambda}{\rm H}({\vec p},{\vec q}+{\textstyle \frac{2}{3}}
  {\vec Q}) ~\right|~ {^3{\rm He}}({\vec p},{\vec q})\right\rangle ~.
\end{eqnarray}
Note that in Eq.~(\ref{eq:efq1}), $\vec{p}$ and $\vec{q}$ denote the momenta of the 
interacting nucleon and the spectators inside the $^3{\rm He}$, 
respectively, whereas $\vec{Q}$ is the
momentum transfer (see Fig.~\ref{fig:hypertriton}).

 The nuclear cross section at forward angles is smaller 
than that of elementary kaon production by two orders of magnitude.
As $\theta_K^{\rm c.m.}$ increases, the cross section drops quickly, 
since the nuclear momentum transfer increases as a function of 
$\theta_K^{\rm c.m.}$. Since this cross section is in fact very small, 
photoproduction of hypertriton provides a great challenge to experimentalist.
Figure \ref{fig:ts_fq2} also shows the significant difference between the cross
sections calculated with the approximation of Eq.~(\ref{eq:approximated_cs}) and 
the full result obtained from Eq.~(\ref{eq:cs_hypertriton}). This discrepancy is   
due to the ``factorization'' approximation, since in the exact  
calculation both spin-independent and spin-dependent amplitudes 
are integrated over the internal momentum and weighted by the two wave 
functions. This also explains why more structures appear in the cross section
of the exact calculation.

\begin{figure}[b]
\centerline{\epsfxsize=9cm \epsffile{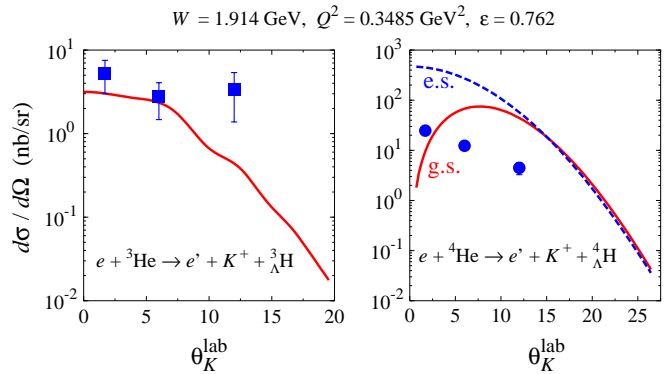}}
\caption{\small Comparison between theoretical calculation 
  and experimental data of the $^3_\Lambda$H (left) and $^4_\Lambda$H (right)
  electroproduction cross sections. 
  Data are taken from \cite{dohrmann}.}
   \label{fig:dohrmann} 
\end{figure}

Very recently, Dohrmann {\it et al.} \cite{dohrmann} measured 
the electroproduction process on both $^3$He and $^4$He targets. 
Although their experiment yielded only three data points in each 
target, this provides the first evidence on the possibility to 
produce hypernuclei via electromagnetic process. Our investigation 
of this process is still in progress. To compare our photoproduction 
results with these new data, we multiply Eq.~(\ref{eq:cs_hypertriton}) 
with the ratio between the electro- and photoproduction cross sections 
of the elementary process. The results for both targets are displayed 
in Fig.~\ref{fig:dohrmann}, where we see that a fair agreement is 
obtained in the case of the hypertriton production and a significant 
discrepancy appears for the $^4$He target. This clearly shows that 
our simple approach urgently requires substantial improvement.

\vspace{2mm}

\section{Moving to Neutron Star}
\subsection{Simple Electron and Neutron Stars}
\begin{figure}[t]
\centerline{\epsfxsize=3.5cm \epsffile{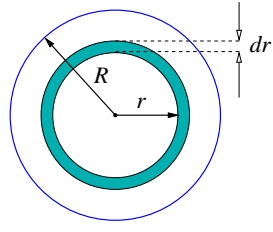}}
\caption{\small A solid sphere with radius $R$ and density $\rho$. The mass of the 
  shell with the thickness $dr$ is $4\pi\rho r^2dr$, whereas the mass of the sphere
  enclosed by this shell is $4\pi\rho r^3/3$.}
   \label{fig:sphere} 
\end{figure}

Let us 
assume that a star has burned all of its ``fuel'' and the gravitational force 
starts to contract the star until nothing but the electron degeneracy pressure tries
to save the star from a gravitational collapse. To calculate the gravitational pressure,
let us consider a solid sphere depicted in Fig.\,\ref{fig:sphere}. From this figure,
it is straightforward to derive the gravitational potential of the shell lying 
between $r$ and $dr$, which in this case can be written as\footnote{For the 
quantum-mechanical derivation of the neutron star radius, we follow 
Gasiorowicz's book \cite{gasiorowicz}.}
\begin{eqnarray}
  \label{eq:grav_potential1}
  d{\cal V}_{\rm grav.} &=& 
  -\frac{G\left(\frac{4}{3}\pi\rho r^3\right)\left(4\pi\rho r^2 dr\right)}{r^2}
  ~=~ -\frac{16\pi^2\rho^2G}{3}\, r^4\, dr~,
\end{eqnarray}
where $\rho$ is the density of the sphere and 
$G$ is the universal gravitational constant. 
By integrating over the entire radius of the sphere we obtain the total potential energy,
\begin{eqnarray}
  \label{eq:grav_potential2}
  {\cal V}_{\rm grav.} &=& -\frac{16\pi^2\rho^2G}{3}\int_0^R r^4\, dr 
  ~=~ -\frac{16\pi^2\rho^2R^5G}{15} ~=~
  -\frac{3}{5}\,GM^2\,\left(\frac{4\pi}{3}\right)^{1/3} V^{-1/3} ~,
\end{eqnarray}
where $M$ is the total mass of the sphere and $V$ is its volume. The gravitational pressure
is obtained by differentiating Eq.\,(\ref{eq:grav_potential2}) to the volume $V$,
\begin{eqnarray}
  \label{eq:grav_pressure}
  p_{\rm grav.} &=& -\frac{\partial{\cal V}_{\rm grav.}}{\partial V}
  ~=~ -\frac{1}{5}\left( \frac{4\pi}{3} \right)^{1/3} G(Nm_n)^2\, V^{-4/3}~,
\end{eqnarray}
where we have assumed that the sphere contains $N$ neutron each with a mass of $m_n$,
i.e., $M=Nm_n$.

 The degeneracy pressure is obtained from Eq.\,(\ref{eq:degenerate_force1}),
which can be written as
\begin{eqnarray}
    \label{eq:degenerate_force2}
    p_{\rm deg.} ~=~ \frac{\hbar^2\pi^3}{15m_e}\left( \frac{3N_e}{\pi}
    \right)^{5/3} V^{-5/3}~,
\end{eqnarray}
with $N_e$ and $m_e$ are the number and the mass of electron.

 If there is an equilibrium between the two forces, then we have 
\begin{eqnarray}
  \frac{1}{5}\left( \frac{4\pi}{3} \right)^{1/3} G(Nm_n)^2\, V^{-4/3} &=&
  \frac{\hbar^2\pi^3}{15m_e} \left( \frac{3N_e}{\pi} \right)^{5/3} V^{-5/3}~,
\end{eqnarray}
where $N$ and $N_e$ are the number of neutron and electron in the star, 
$m_n$ and $m_e$ are the neutron and electron masses, respectively.
Solving for the star radius, we obtain 
\begin{eqnarray}
  \label{eq:electron_star_radius}
  R &=&  \left(\frac{81\pi^2}{128} \right)^{1/3} \frac{\hbar^2\pi^3}{Gm_em_n^2}\, N^{-1/3}~.
\end{eqnarray}
For a star of one solar mass, we have $N\approx 1.2\times 10^{57}$
and assuming $N_e=N/2$ we obtain
\begin{eqnarray}
  R_{\rm es} ~\approx ~  1.1\times 10^4 ~{\rm km} ~.
\end{eqnarray}
Such a star, for which its collapse is halted by electron degeneracy, is 
called white dwarf.

 However, if we started with a heavier star then 
the gravitational force would be too strong and the electron degeneracy force 
could not overcome it. As a consequence, electrons and protons would start the process 
\begin{eqnarray}
  \label{eq:urca}
  e^- +\, p ~\to~ \nu_e +\, n~.
\end{eqnarray}
This process indicates that all atoms in the star start to collapse.
Since the neutrino interacts very seldom with other particles, it
will exit the star and we are left with a {\sl pure neutron star}. 
To calculate the radius of this star we can
use Eq.\,(\ref{eq:electron_star_radius}) and substitute the mass of
electron with the mass of neutron, i.e.,
\begin{eqnarray}
  \label{eq:neutron_star_radius}
  R_{\rm ns} &=&  \left(\frac{81\pi^2}{128} \right)^{1/3} \frac{\hbar^2\pi^3}{Gm_n^3}\, N^{-1/3}~.
\end{eqnarray}
If we started with a star with mass equals to the solar mass,\footnote{Only for example.
This is given in order to be in line with the discussion of Gasiorowicz.
A more consistent calculation indicates that this could be true only if
we started with a star with the mass greater than 1.4 of the solar mass.}
we will then end up with a
neutron star with $R_{\rm ns}\approx 10$ km. This result should be
compared with the solar radius which is approximately $7\times 10^8$ km.

\subsection{A Slightly More Realistic Neutron Star}
The neutron star described in the previous subsection is clearly far from
realistic. First of all, because
a real neutron star contains not only neutron. Why? Let us consider again
the process shown in Eq.\,(\ref{eq:urca}). Given that a free neutron
will decay in about 15 minutes via
\begin{eqnarray}
  \label{eq:urca1}
  n ~\to~ p~+~e^-+~\bar{\nu}_e~,
\end{eqnarray}
certain amount of the protons should be present in the neutron star to 
prevent this weak decay. Most of the energies in the reaction of
Eq.\,(\ref{eq:urca1}) is carried away by the anti-neutrino and, therefore,
the protons can only occupy the low-energy levels. Once all these levels
are filled, Pauli exclusion principle prevents the decay from
taking place. On the other hand, the appearance of these protons must be 
accompanied by the presence of the same amount of electrons, 
in order to make the charge of the star neutral. 

\begin{figure}[htb]
\centerline{\epsfxsize=10cm \epsffile{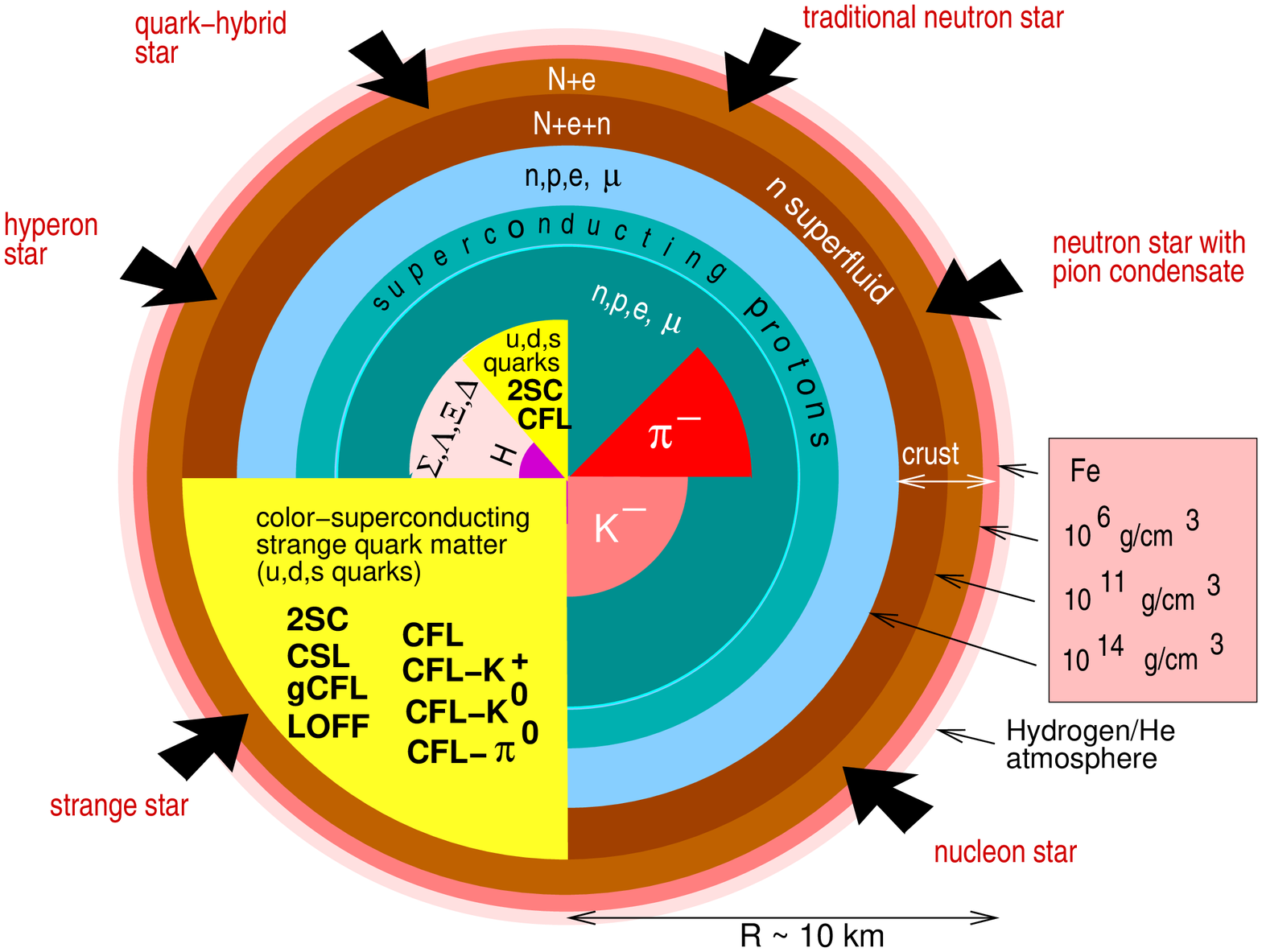}}
\caption{\small The structure and phase of subatomic matter inside the neutron star as
   predicted by theory. Also shown in this figure, the density of
   the crust layers \cite{Weber:2004kj}
   (Courtesy of Prof. Fridolin Weber, San 
   Diego State University).}
   \label{fig:fweber} 
\end{figure}

Inside the neutron star the pressure is increasing. As a consequence, other particles 
such as $\mu, \pi$, and $K$ mesons may appear. Going deeper and deeper we will
meet exotic particles, such as $\Lambda, \Sigma, \Xi$, and $\Delta$, or even
the up, down, and strange quarks. This can be nicely seen in Fig.\,\ref{fig:fweber}.

 Besides that, since the pressure is extremely strong, relativistic corrections start to
play a significant role. To understand this correction, let us go back to the
solid sphere shown in Fig.\,\ref{fig:sphere}. The gravitational force between
the mass of the shell ($dm$) and the mass enclosed by it ($m$) can be written as
\begin{eqnarray}
  \label{eq:force_m_dm}
  dF &=& -\,\frac{G\, m\, dm}{r^2}
  ~=~ -\,\frac{G\, m(r)\, \rho(r)\, A(r)\, dr}{r^2} ~,
\end{eqnarray}
where $\rho(r)$ is the density of the shell and 
$A(r)$ is the area covered by the shell. By using $dp=dF/A$, we obtain
\begin{eqnarray}
  \label{eq:pressure_newtonian_star}
  \frac{dp}{dr} &=& -\,\frac{G\,\rho(r)\, m(r)}{r^2}~.
\end{eqnarray}
We also note that the mass $m(r)$ in Eq.\,(\ref{eq:pressure_newtonian_star})
is related to the density $\rho(r)$ by
\begin{eqnarray}
  \label{eq:mass_density}
  \frac{dm}{dr} &=& 4\pi\,r^2\rho(r) ~.
\end{eqnarray}
Equations (\ref{eq:pressure_newtonian_star}) and (\ref{eq:mass_density}) provide
the tools to investigate a Newtonian star, i.e., a star for which the mass is not so
large and therefore no relativistic correction is needed. 
These equations can be integrated self consistently, from 
$r=0$ (where $p=p_0$ and $\rho=\rho_{\rm c}$) up to $r=R$ (where $p=0$), to obtain
the pressure $p(r)$ and, therefore, the radius $R$ and the mass $M$ of the star.
To perform 
this integration, one needs the information of the matter density $\rho(r)$
in terms of the pressure $p(r)$. This relationship is well known as the
equation of state (EOS) of the matter building up the star.

 There is a certain class of stars for which the corresponding 
EOS is quite simple and takes the form
\begin{eqnarray}
  \label{eq:polytrope}
  p&=& K\,\rho^\gamma ~,
\end{eqnarray}
where $K$ is the proportionality constant which depends on the entropy per nucleon
and chemical composition. 
Any star fulfilling Eq.\,(\ref{eq:polytrope}) is called a {\sl polytrope}. 
The case of {\sl polytrope} is of interest since there are analytic
solutions in terms of Lane-Emden function $\theta(\xi)$ \cite{weinberg}.
In this case the radius of the star is given by
\begin{eqnarray}
  \label{eq:radius_polytrope}
  R&=& \left[ \frac{K\gamma}{4\pi G(\gamma -1)} \right]^{1/2}\rho_{\rm c}^{(\gamma-2)/2}\xi_1~,
\end{eqnarray}
whereas its mass is given by
\begin{eqnarray}
  \label{eq:mass_polytrope}
  M &=& 4\pi\rho_{\rm c}^{(3\gamma-4)/2}\left[ \frac{K\gamma}{4\pi G(\gamma -1)} \right]^{3/2}
  \xi_1^2\,\left|\theta'(\xi_1)\right| ~,
\end{eqnarray}
where both $\xi_1$ and $\theta'(\xi_1)$ are functions of the polytropic index $\gamma$
and tabulated in Ref.\,\cite{weinberg}.
By eliminating $\rho_{\rm c}$ in Eqs.\,(\ref{eq:radius_polytrope}) and
(\ref{eq:mass_polytrope}), it can be shown that the relation between mass and radius 
of this type of star is independent of the choice of $\rho_{\rm c}$.

Most of the known stars 
can be sufficiently described as Newtonian ones. Besides that, the 
compact stars known as white dwarfs also fall into this category. 

\begin{figure}[t]
\centerline{\epsfxsize=8cm \epsffile{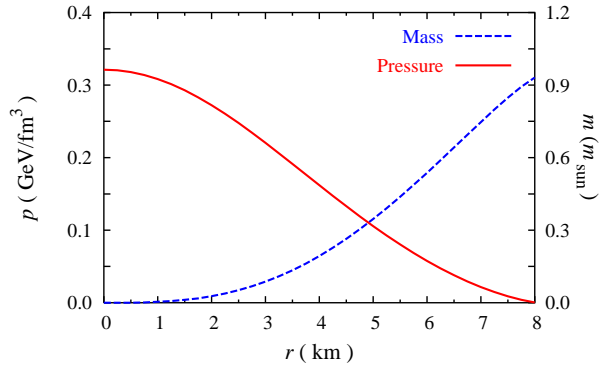}}
\caption{\small The pressure and mass of a neutron star calculated from
  Eqs.\,(\ref{eq:pressure_neutron_star}) and (\ref{eq:mass_density}) as a
  function of the distance from its center. In obtaining this result
  we have used $\rho_{\rm c}=5\rho_0$, where $\rho_0=0.17~{\rm fm}^{-3}$, 
  the normal nuclear density \cite{nyiri}.}
   \label{fig:pressure} 
\end{figure}

 For more massive stars, such as neutron stars, the relativistic correction is needed.
This can be performed by multiplying Eq.\,(\ref{eq:pressure_newtonian_star}) with
three correction factors, 
\begin{eqnarray}
  \label{eq:pressure_neutron_star}
  \frac{dp}{dr} &=& -\,\frac{G\,\rho(r)\, m(r)}{r^2}
  \underbrace{\left[1+\frac{p(r)}{\rho(r)c^2}\right]
  \left[1+\frac{4\pi r^3p(r)}{m(r)c^2}\right]\left[1-
  \frac{2Gm(r)}{rc^2}\right]}_{\rm relativistic~corrections}~,
\end{eqnarray}
while maintaining the differential equation for $m(r)$ 
[Eq.\,(\ref{eq:mass_density})]. Equation (\ref{eq:pressure_neutron_star})
is known as the Tolman-Oppenheimer-Volkoff (TOV) equation \cite{Tolman:1939jz}.
In this equation the first two factors in square brackets
are due to special relativity corrections of order $v^2/c^2$. It is 
obvious that these factors go to zero in the limit of $c\to\infty$.
The third factor is a general relativity correction. It is also
clear that this factor can be neglected in the non-relativistic
case. It is obvious from Eq.\,(\ref{eq:pressure_neutron_star}) that
an analytic solution of the TOV equation is far from possible.
The discussion of the solution to this equation by using an advanced
EOS is clearly beyond the scope of this paper. The interested reader
can consult Ref.\,\cite{Lattimer:2006xb} (as well as references therein) 
for the latest review on this topic. As an example of the numerical solution
to the coupled equations (\ref{eq:mass_density}) and (\ref{eq:pressure_neutron_star})
we show the pressure and the mass of neutron star as a function of the distance
from the center of the star in Fig.\,\ref{fig:pressure}.
Here, a simple form of the energy density of nuclear matter consisting a compressional
term, a symmetry term, and binding energy term, has been used for the corresponding 
EOS \cite{nyiri}.

\subsection{Neutrino Mean Free Path in Neutron Stars}
In discussing the formation of a neutron star, i.e., Eqs.\,(\ref{eq:urca}) 
and (\ref{eq:urca1}), we understand that both neutrino and anti-neutrino
immediately leave the neutron star as soon as they have been created. 
Moreover, it is also comprehended that
most of the energies coming from the two reactions are carried away by neutrino
and anti-neutrino. This has an important consequence in the cooling process
of a neutron star, i.e., whereas a conventional body dissipates its energy 
through photon emissions (electromagnetic radiations), a neutron star cools 
down via emitting neutrinos and anti-neutrinos.

 Therefore, understanding of the interaction between neutrino (as well as 
anti-neutrino, from now on we will only mention neutrino for the sake of
brevity) and neutron star is crucial in the investigation of the 
neutron star formation. Since the neutron star density is extremely high
(see Fig.\,\ref{fig:fweber}), such interaction is more probable than
in the conventional matter. In fact, the neutrino mean free path (the
distance between two interactions, from now on called NMFP) 
could in certain cases be smaller 
than the diameter of the neutron star. If this happened, we could 
say that the neutrino is trapped inside the neutron star. The ``neutrino
trapping'' can have a serious consequence in the properties of neutron
stars. Indeed, it plays an important role in the supernova explosion
or the neutron star cooling process.

 Although in the Standard Model massless neutrinos have zero magnetic 
moment and electronic charge, there are some evidences that 
neutrino-electron has a magnetic moment ($\mu_{\nu}$) 
smaller than $1.0 \times 10^{-10} \mu_B$ at 90\% confidence 
level~\cite{munu}, where $\mu_B$ is the Bohr 
magneton. A stronger bound  of $\mu_{\nu} \leq 3.0 
\times 10^{-12} \mu_B$ also exists from astrophysical 
observation, particularly from the study of the red giant 
population in globular clusters \cite{Raffelt99}. 
On the other hand the neutrino charge radius of $\nu_e$ 
has been estimated from the LAMF experiment \cite{vilain} 
to be $R^2=(0.9 \pm 2.7)\times 10^{-32}~{\rm cm}^{2}=
(22.5 \pm 67.5)\times 10^{-12}~{\rm MeV}^{-2}$, 
while the plasmon decay in the globular cluster 
star predicts the limit of $e_{\nu} \leq 2 \times 10^{-14}\,e$ \cite{Raffelt99}, 
where $e$ is electron charge. 

\begin{figure}[t]
~~~
\begin{minipage}[h]{8cm}
{\epsfxsize=8cm \epsffile{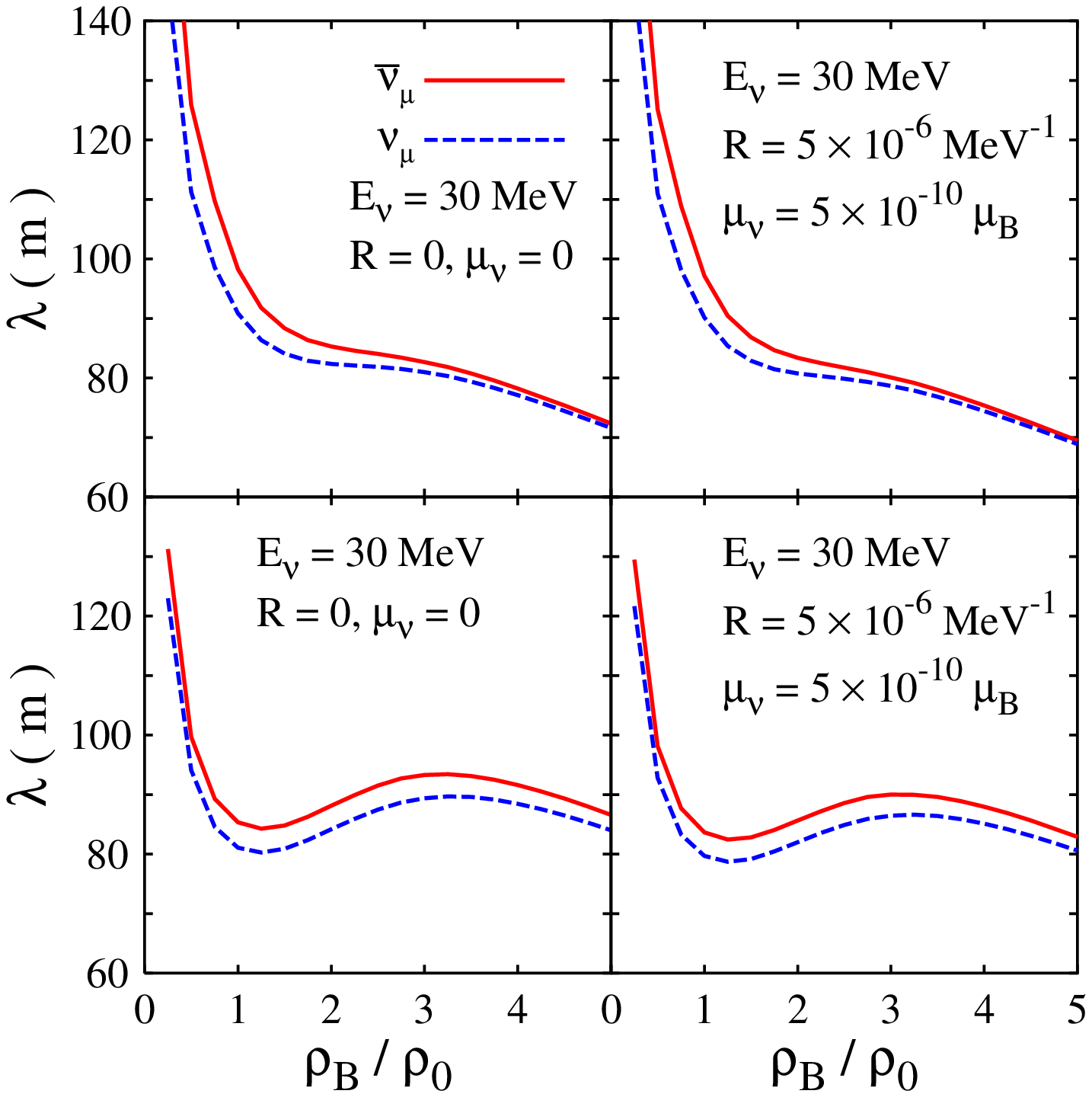}}
\end{minipage}
%
%
\begin{minipage}[h]{8cm}
{\epsfxsize=8cm \epsffile{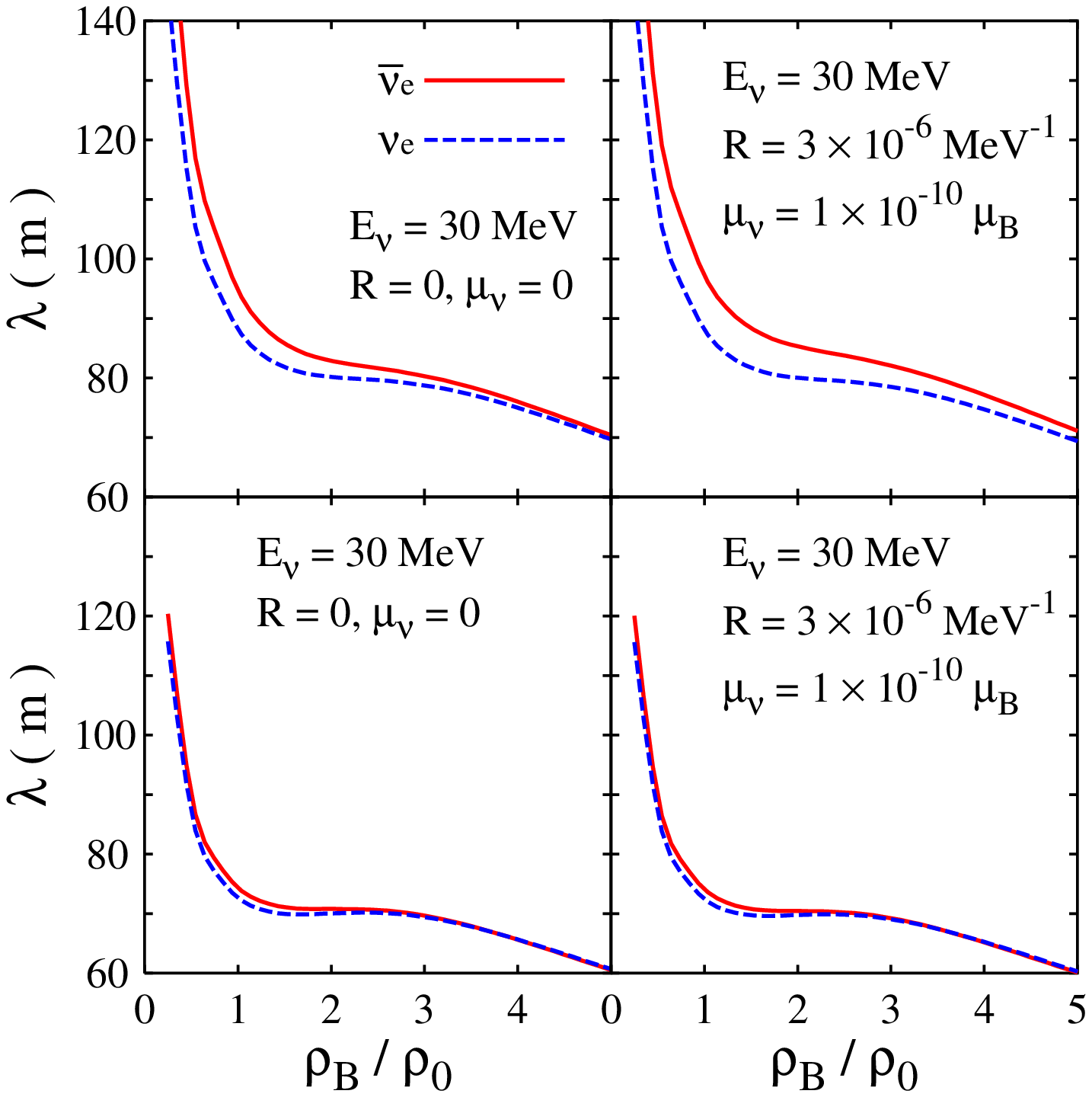}}
\end{minipage}
\caption{\small {\bf (Left)} The muon-neutrino and muon-antineutrino mean free paths as a function 
    of the ratio between nucleon and nuclear saturation densities. Results 
    for neutrinoless matter are shown in upper panels, whereas results for 
    neutrino trapping with $Y_{l_e}$= $0.3$ are shown in lower panels.
    {\bf (Right)} Same as in the left panel, but for electron-neutrino and 
    electron-antineutrino \cite{Hutauruk:2006re}.}
  \label{fig:neutrino_mfp}
\end{figure}

 The NMFP (symbolized by ${\lambda}$) is obtained by integrating the cross section 
of the neutrino-dense-matter interaction 
over the time- and vector-component of the neutrino momentum transfer,
\begin{eqnarray}
\label{eq:nmfp}
\frac{1}{{\lambda}(E_{{\nu}})} = \int_{q_{0}}^{2E_{{\nu}}-q_{0}}d|{\vec{q}}|
\int_{0}^{2E_{{\nu}}}dq_{0}\frac{2{\pi}|{\vec{q}}|}{E'_{{\nu}}E_{{\nu}}}~
\frac{1}{V}\frac{d^3{\sigma}}{d^2{\Omega}'dE'_{{\nu}}}~,
\end{eqnarray}
where the cross section for every type of neutrino can be calculated from
\begin{eqnarray}
\frac{1}{V}\frac{d^{3} \sigma}{d^{2}{\Omega}'dE'_{{\nu}}} &=& -\frac{G_{F}}{32{\pi}^{2}}\frac{E'_{{\nu}}}{E_{{\nu}}}{\rm Im}(L_{{\mu}{\nu}}{\Pi}^{{\mu}{\nu}}). 
\label{eq:cross}
\end{eqnarray}
where $L_{\mu\nu}$ is the neutrino tensor and ${\Pi}^{{\mu}{\nu}}$ 
is the target polarization tensor, which defines the
particle species. Details of the procedure for calculating Eqs.
(\ref{eq:nmfp}) and (\ref{eq:cross}) can be found in Ref.\,\cite{parada04}

In order to see the effect more clearly, in this calculation we 
use relatively large values of neutrino dipole moments and charge 
radii. Furthermore, besides using conventional matter, 
we also consider matter with neutrino trapping. The existence 
of the neutrino in matter allows for the presence of a relatively 
large number of protons and electrons compared to the case of 
neutrinoless matter. The appearance of these constituents is 
then followed by the appearance  of  a small number of muons 
at a density larger than two times the nuclear saturation density.

 The results for muon-neutrino and electron-neutrino mean free 
paths are shown in Fig.~\ref{fig:neutrino_mfp}, where we chose 
the neutrino energy $E_\nu=30$ MeV. From this figure it is obvious 
that in the case of neutrinoless matter,  contribution from 
the neutrino charge radius is too small compared to that 
from the nucleon weak magnetism. On the other hand, in the 
case of neutrino trapping, but with  zero muon-neutrino 
dipole moment and charge radius, the difference in the 
neutrino and antineutrino mean free paths is significant, 
especially at high densities. Although almost similar to 
the neutrinoless matter case, contribution from the neutrino 
charge radius is very small, and therefore it is invisible 
in this kinematics. It is also apparent that in both cases 
for densities around $(2-3)\rho_0$, $\lambda_{\nu}$ and 
$\lambda_{\bar{\nu}}$ behave differently, i.e., if 
neutrinos are present in matter, the mean free paths 
increase the density, but if  neutrinos are absent, 
the opposite phenomenon is observed. We have shown that
this result is more pronounced at higher neutrino energies
\cite{Hutauruk:2006re}.

\begin{figure}[t]
\centerline{\epsfxsize=12cm \epsffile{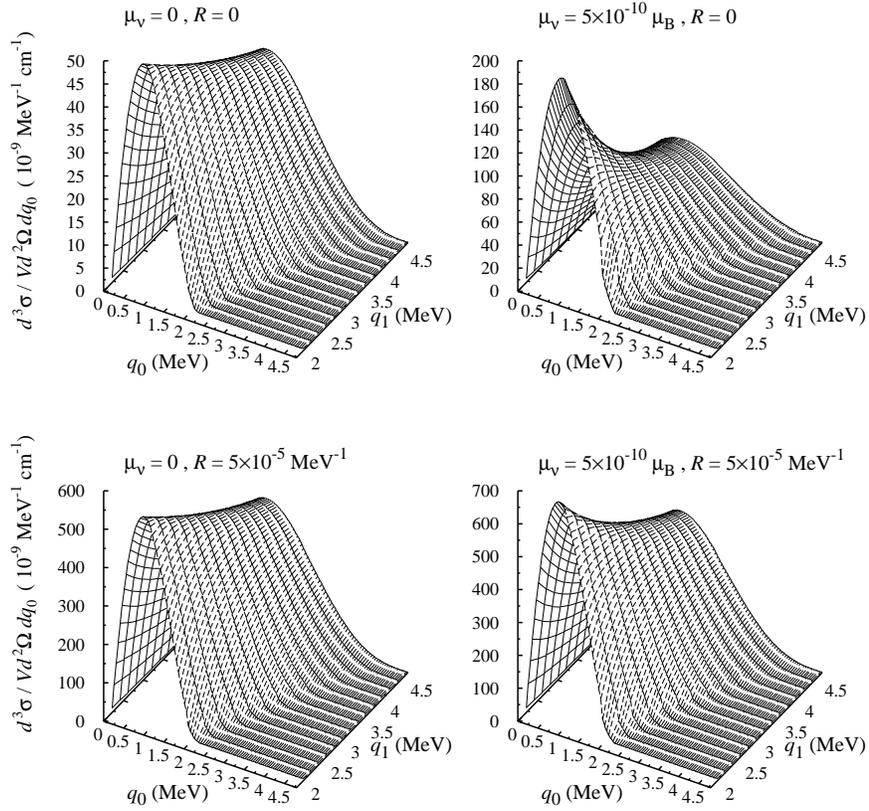}}
\caption{Total differential cross sections as functions of the 
  momentum transfer $q^\mu=(q_0,\vec{q}_1)$ 
  for the case where neutrinos are trapped with $Y_{l_e} = 0.3$ and
  baryon density $\rho_B=5\rho_0$, while $E_{\nu}=5$ MeV
  \cite{Sulaksono:2006eu}.}
   \label{fig:cs_3dim} 
\end{figure}

 In the case of electron-neutrino, although the effect is 
less significant for neutrinoless matter at high densities, 
contribution from the neutrino charge radius yields an 
enhancement to the difference between $\lambda_{\nu}$ 
and $\lambda_{\bar{\nu}}$. On the contrary, for the case 
of zero neutrino dipole moment and charge radius,  but 
with neutrino trapping, the difference of neutrino and 
antineutrino mean free paths is suppressed. Thus for 
this kind of matter, we may conclude that the neutrino 
charge radius does not contribute to the mean free path 
difference.

Another alternative to study the properties of the 
neutrino-dense matter interaction is by investigating 
the sensitivity of the corresponding differential 
cross section to these possibly nonzero neutrino 
electromagnetic properties. 
The result for the case of neutrino trapping with
leptonic fraction $Y_{l_e}=0.3$ and baryon density 
$\rho_B=5\rho_0$ is shown in Fig.~\ref{fig:cs_3dim}, where  we plot 
the total differential cross sections as functions 
of the time- and vector-component of the four-momentum 
transfer $q^\mu$.

From Fig.~\ref{fig:cs_3dim} we can clearly see that
the cross section magnitude is quite sensitive
to the neutrino electromagnetic properties.
Including contribution from the neutrino magnetic moment
and charge radius leads to a cross section which is 14
times larger than that of point neutrino. It is also
apparent that most of this contribution come from
the neutrino charge radius, whereas the different shape
in the $q_1$ distribution is mainly driven by the neutrino
magnetic moment.

We have also compared this result to the case of neutrinoless
matter \cite{Sulaksono:2006eu}. We found that, in general,  
neutrino trapping enhances the magnitude of the cross section, 
as we expected, since matter with neutrino trapping contains
more protons and, hence, electromagnetic interactions are 
more likely to occur. If we switched off the neutrino
electromagnetic properties, this behavior clearly
disappears. 

\subsection{Critical Density of Dense Matter}
From Fig.~\ref{fig:fweber} it is apparent that a 
neutron star is expected to have a solid inner crust 
above its liquid mantle. The mass of this crust has
been known to be sensitively dependent on the density 
of the inner edge and on the EOS. Therefore, 
investigation of the dynamical instability of 
dense matter will certainly shed important information 
on the inner structure of neutron stars.

\begin{figure}[t]
\begin{minipage}[b]{8cm}
{\epsfxsize=8cm \epsffile{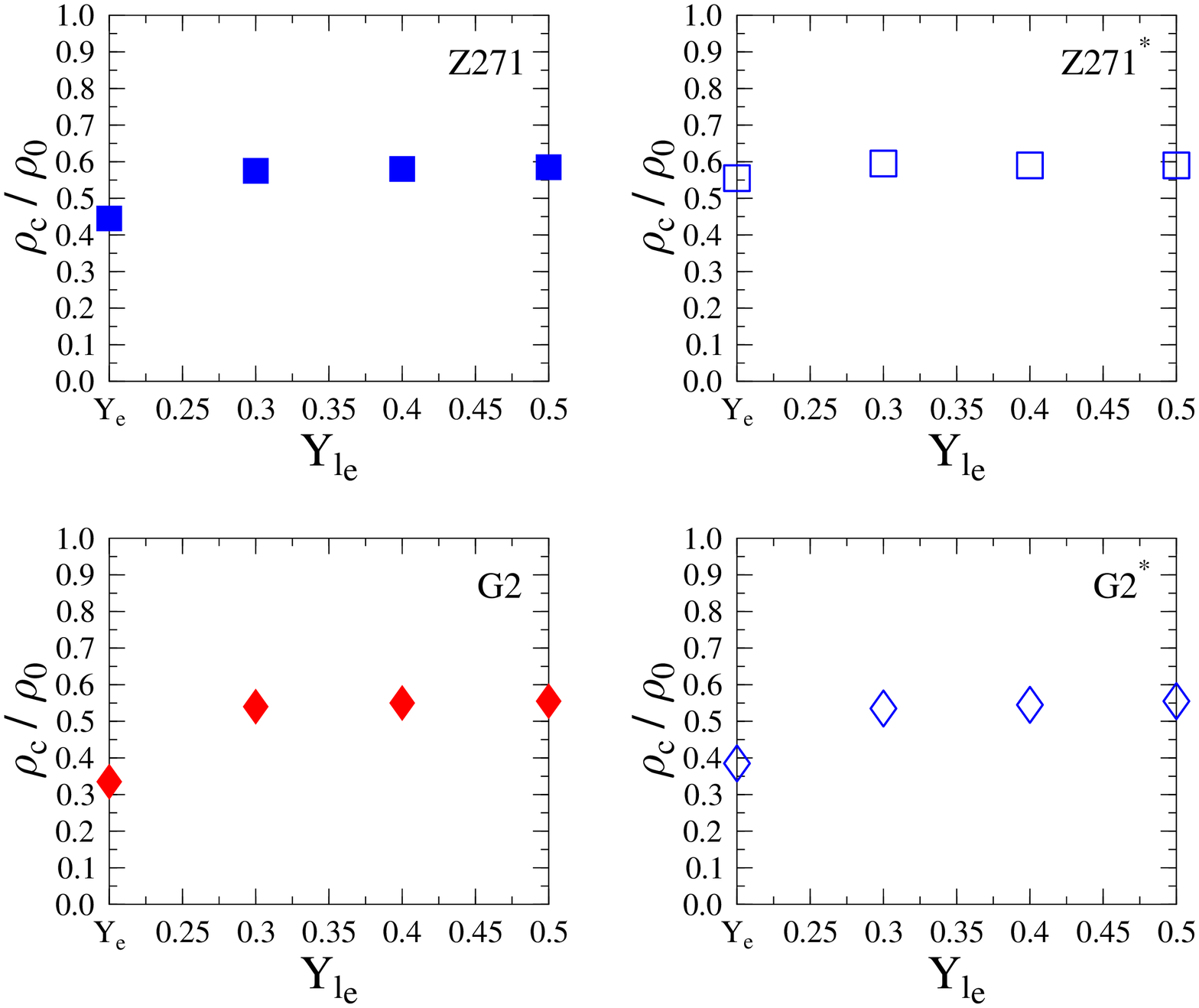}}
\end{minipage}
\hspace{\fill}
\begin{minipage}[b]{8cm}
{\epsfxsize=8cm \epsffile{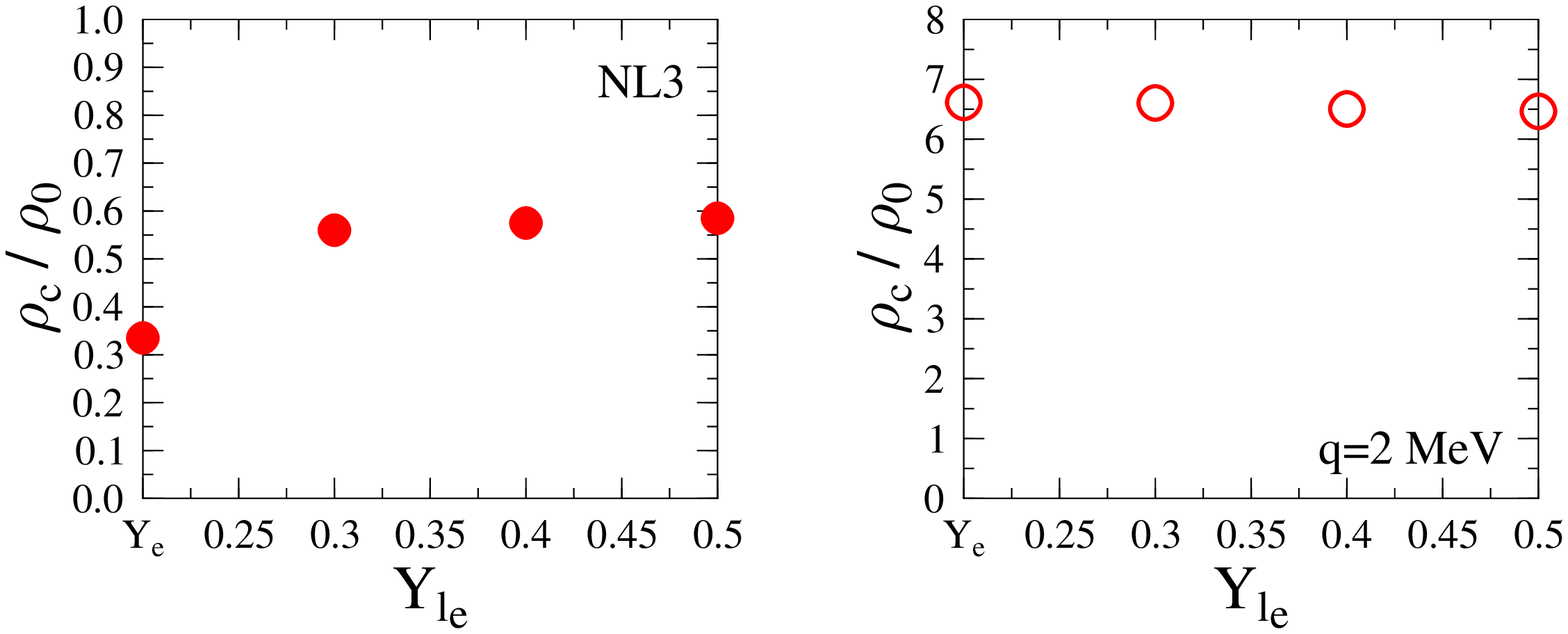}}
\end{minipage}
\caption{Critical densities of different RMF 
  models as a function of the neutrino fraction in matter.}
  \label{fig:instability}
\end{figure}

Recent investigations on this topic have disclosed that
the critical density ($\rho_c$), a density at which the
uniform liquid becomes unstable to a small density
fluctuation, can be used as a good approximation of 
the edge density of the crust \cite{edge_density}. 
The important aspects of their finding are 
that the inner edge of the crust density to be  
$\rho_{\rm edge} = 0.08 ~{\rm fm^{-3}}$ and a measurement 
of the neutron radius in the $^{208}{\rm Pb}$ nucleus 
will provide useful information on the $\rho_c$. 

To this end we have investigated the behavior of 
the predicted critical density ($\rho_c$) in matter 
with and without neutrino trapping for several
different relativistic mean field (RMF) models 
\cite{Sulaksono:2006fb}. The ratio between the 
critical density and the normal nuclear density 
as a function of the neutrino fraction in matter is 
shown in Fig.~\ref{fig:instability} for several 
different RMF models.

It is found that different 
treatments in the isovector-vector sector of 
RMF models yield more 
substantial effects in matter without neutrino 
trapping than in matter with neutrino trapping. 
Moreover, for matter with neutrino trapping, 
the value of $\rho_c$ does not significantly 
change with the variation of the models nor 
with the variation of the neutrino fraction 
in matter. In this case, the value of $\rho_c$ 
is larger for matter with neutrino trapping. 
These effects are due to the interplay between 
the major role of matter composition and the 
role of the effective masses of mesons and nucleons. 

It is also found that the additional nonlinear terms 
of Horowitz-Piekarewicz and Effective RMF models 
can prevent another instability at relatively high 
densities from appearing. This can be traced back 
to the effective $\sigma$ mass, which goes to zero 
when the density approaches $6.5\rho_0$.

\section{Connecting What We Have Learned So Far}
So far, we have discussed hypernucleus and neutron star
starting from the Pauli exclusion principle. We know
that the important ingredient of a hypernucleus is
hyperon, whereas the basic building block of a neutron
star is neutron. According to their symmetry, both 
neutron and hyperon are classified in the SU(3) 
baryon octet. We have also implicitly built an argument 
that without this principle we know nothing about the 
physics of hypernuclear and neutron star.

\begin{figure}[t]
\centerline{\epsfxsize=8cm \epsffile{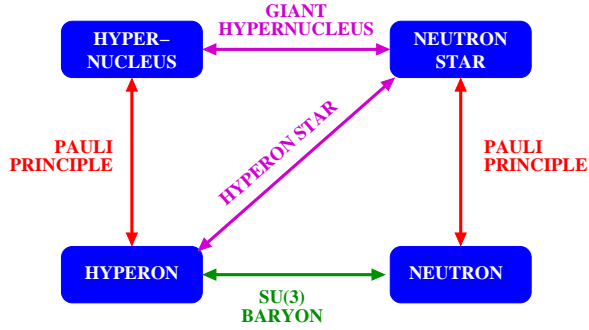}}
\caption{Connecting and extending from what we have learned so far.}
   \label{fig:relasi} 
\end{figure}

Now, we can nicely connect hyperon, neutron, hypernucleus,
and neutron star as shown in Fig.~\ref{fig:relasi}. Starting
from this figure, we may ask: what can be constructed based 
on our knowledge about them? A tentative answer is also
shown in this figure. 

Within the RMF model, the question whether a neutron star 
could be considered as a giant hypernucleus has been asked 
more than 20 years ago \cite{Glendenning:1984jr}. Earlier
discussion which showed that the existence of hyperon in 
neutron star is plausible has been done based on Fermi 
gas model \cite{ambartsumyan}. For our current discussion 
this would not be too 
difficult to understand if we look back at Fig.~\ref{fig:fweber},
where we can clearly see that at certain density (depth)
there is a $\Lambda$ population. In fact, Glendenning 
has found that this population starts to grow at baryon
density $\rho_B\approx 0.4~{\rm fm}^{-1}$ and 
starts to dominate all star ingredients at 
$\rho_B\approx 0.9~{\rm fm}^{-1}$ \cite{Glendenning:1984jr}. 
Such population is found in the cores of the relatively 
heavier neutron stars and is about 15\%--20\% of the
total baryon population.

We can continue the discussion to the hyperon stars
as well as quark stars. They are indeed more massive,
more complicated, and more interesting than the neutron
star. However, we have to stop this 
discussion here, otherwise I have the feeling that
our experimentalist colleagues would complain
that we, theoreticians, are too imaginative
and too creative. Could this be true?
The next section could 
convince you that this could be true.

\section{Econophysics, are Theoreticians too Creative?}
\subsection{Why Econophysics?}
 
The scientific challenge to understand the nature of 
complex systems is too tempting to the physicists. 
Some of them have been fascinated by the 
huge, and also growing, amount of economics data recorded minutes 
by minutes for decades. Among these interesting data, the fluctuation 
of stock exchange indices is of special interest, since it  might 
indirectly reflect the economic situation in a certain region. 
Furthermore, the advancement in computing capabilities has enabled us 
to handle this large amount of data, unlike the situation almost 40 years ago, 
when Mandelbrot investigated approximately 2000 data points of cotton
prices \cite{mandelbrot}.

It is said that such studies could explain the nature of
interacting elements in the complex systems and, therefore, could 
help to forecast economic fluctuations in the future. In other 
words, these studies were intended to produce new results in 
economics, which might help us to avoid economic ``earthquakes'' 
such as what happened in Indonesia several years ago \cite{stanley2000}. 

\subsection{Time Series Analysis}
We start our analysis with an investigation of the daily index 
returns, which are defined as
\begin{eqnarray}
  \label{eq:return}
  Z_{\Delta t}(t) &=& \ln Y(t+\Delta t) - \ln Y(t) 
  ~=~ \ln\, [Y(t+\Delta t)/Y(t)] ,
\end{eqnarray}
where $Y(t)$ indicates the closing index of the stock at day $t$. We use the
time series data from two different stock indices, the Jakarta Stock 
Exchange Index (abbreviated with IHSG, an acronym of {\it Indeks Harga
Saham Gabungan} or the composite stock exchange price index) and the 
Kuala Lumpur Stock Exchange index (KLSE), which belong to different countries. 
Comparing 
the two indices would be very interesting since both Indonesia and Malaysia 
underwent the same monetary crisis in 1997, which was then followed by 
financial crashes in almost all economic sectors, but with quite different 
economic situations. 

\begin{figure}[htb]
\begin{minipage}[h]{8cm}
{\epsfxsize=8cm \epsffile{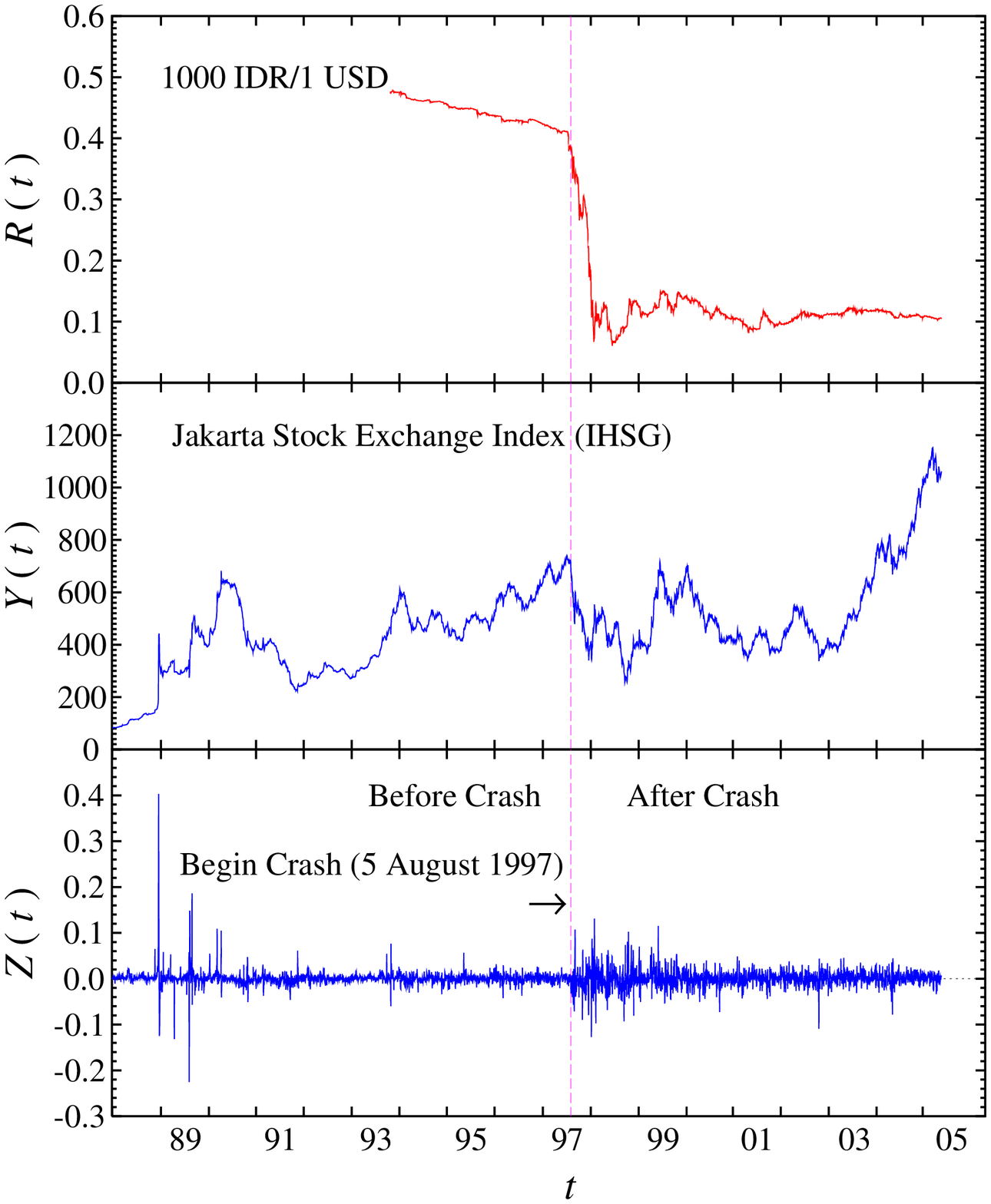}}
\end{minipage}
\hspace{\fill}
\begin{minipage}[h]{8cm}
{\epsfxsize=8cm \epsffile{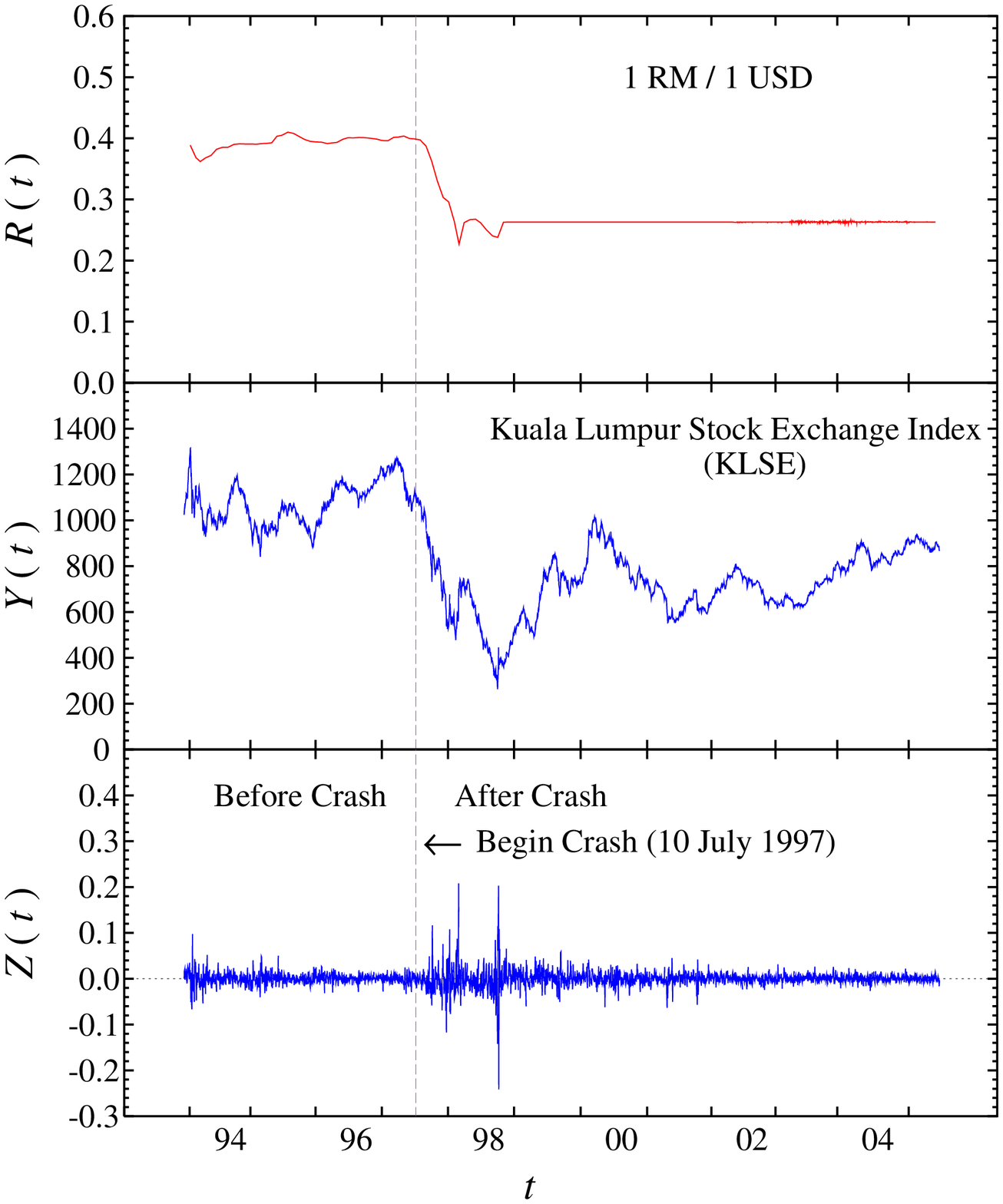}}
\end{minipage}
\caption{\small {\bf (Left)} The ratio between 1000 Indonesian Rupiah and 1 US Dollar
        (top), the Jakarta Stock Exchange Index (IHSG) $Y(t)$ (middle) and
        the logarithmic return $Z(t)$ calculated from Eq.\,(\ref{eq:return})
        as a function of time $t$ sampled with 
        $\Delta t=1$ day (bottom). 
        {\bf (Right)} The ratio between 1 Malaysian Ringgit and 1 US Dollar (top), 
        the Kuala Lumpur Stock Exchange Index (KLSE) $Y(t)$ (middle) and
        the logarithmic return $Z(t)$ (bottom) as a function of time
        sampled for $\Delta t=1$ day. On 1st September 1998 Malaysia 
        imposed currency controls, the Ringgit was pegged with US Dollar 
        with a fixed rate of 3.80 Ringgit per Dollar. 
        The arrows in the bottom panels
        indicate the time position when the IHSG and KLSE started to crash.
        Data are taken from Ref. \protect\cite{oanda} (IDR/USD and RM/USD exchange rates) 
        Ref. \protect\cite{bej} (IHSG stock index), and Ref.
        \protect\cite{yfinance}  (KLSE stock index).}
   \label{fig:returns}
\end{figure}

 Figure \ref{fig:returns} shows the time series of the IHSG and KLSE indices 
along with their logarithmic returns calculated by using Eq.\,(\ref{eq:return}). 
For comparison, in this figure (top panels) we also display the historical time series
of the exchange rate of Indonesian Rupiah and Malaysian Ringgit, since
the economic crashes started with the devastating decline of these rates.
It is obvious from this figure that the fluctuation $Y(t)$ of both indices 
is more dramatic than that of S\&P 500, indicating that in this case the 
situation is far more complex. 
After the crash the magnitude of returns $Z(t)$ is significantly larger 
in both indices, or, in the economics language, the probability to gain 
or to loose becomes larger than before.

\begin{figure}[!]
\centerline{\epsfxsize=8.5cm \epsffile{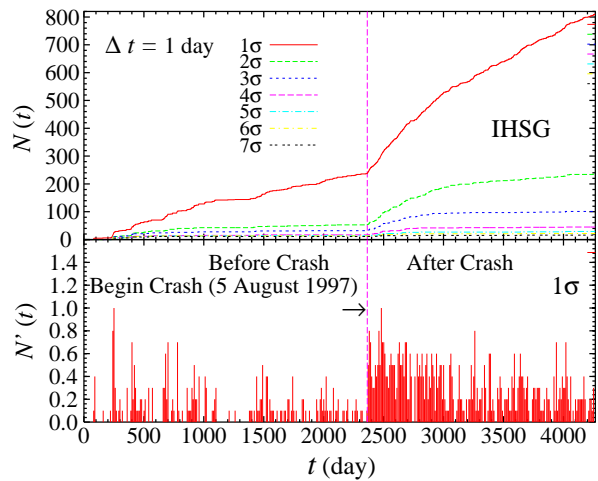}}
\caption{\small Number of the returns given by Eq.\,(\ref{eq:return}) that 
  exceeds one up to seven standard deviations ($\sigma$) for the IHSG (top 
  panel) and the corresponding derivatives at $1\sigma$ (lower panel).}
   \label{fig:omori} 
\end{figure}

\subsection{Omori Law, from Geophysics to Finance Physics}
 To accurately determine the time position of the crash we make use of the Omori law. 
Omori law, which was originally used in geophysics \cite{omori1894}, states that the
number of aftershock earthquakes per unit time measured at time $t$ after the main
earthquake decays as a power law. Practically it is written as
\begin{eqnarray}
  \label{eq:omori1}
  n(t)&=& K\,(t+\tau)^{-p}~,
\end{eqnarray}
where $K$ and $\tau$ are two positive constants. The cumulative number of aftershocks
is obtained by integrating Eq.\,(\ref{eq:omori1}) from 0 to $t$, i.e.,
\begin{eqnarray}
  \label{eq:omori2}
  N(t) &=& \int_0^t dt'\, K(t'+\tau)^{-p}
  ~=~ \frac{K\left[(t+\tau)^{1-p}-\tau^{1-p}\right]}{1-p}~.
\end{eqnarray}
This number is found to be more relevant to compare with the presently analyzed data.
By calculating the number of returns given by Eq.\,(\ref{eq:return}) that exceeds one
standard deviation ($\sigma$) up to $7\sigma$, we arrive at Fig.\,\ref{fig:omori}
for the IHSG case.
From this figure we can clearly see that there exists a discontinuity on 5 August 1997,
which indicates the start of a crash process. To exactly find the position
of this discontinuity we calculate the first derivative of $N(t)$ for $N(t)$ obtained with
$1\sigma$, shown in the lower panels of Fig.\,\ref{fig:omori}.
From these three panels we are convinced that the IHSG started its  crash process on
5 August 1997. The same procedure has been also applied to the KLSE index, from which we 
obtain that the KLSE started to crash on 10 July 1997, almost one month before the IHSG case.
The crash positions of the two indices are indicated by the vertical lines in 
Fig.\,\ref{fig:returns}.

\begin{figure}[t]
\centerline{\epsfxsize=16cm \epsffile{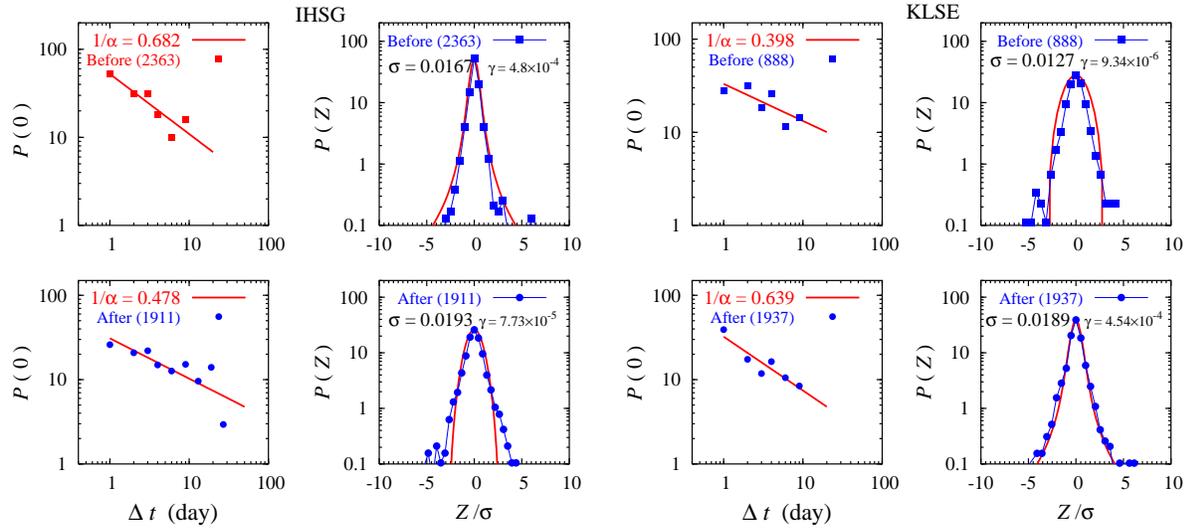}}
\caption{\small {\bf (Left)} Log-log plots of the IHSG probability of return
      to the origin as a function of the sampling time $\Delta t$
      before and after crash. Solid lines are obtained from 
      linear regressions to this probability. The slopes of
      these lines, which are equal to negative inverse of the L\'evy
      stable distribution indices, are shown in the left panels.
      Comparison between normalized distribution 
      functions for $\Delta t=1$ and normalized L\'evy stable distributions
      with parameters obtained from the left figures is shown in the second
      column panels. The standard deviation 
      and the number of data used are also shown in these panels.
      {\bf (Right)} Same as in the four left panels, but for the KLSE case 
      before and after crash.}
   \label{fig:distxyf0}
\end{figure}
\begin{figure}[!t]
\centerline{\epsfxsize=8cm \epsffile{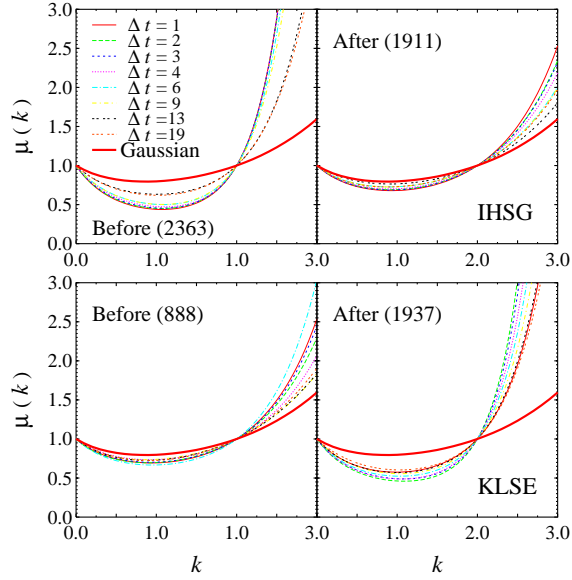}}
\caption{\small Moments of the distribution of the IHSG (top) and KLSE (bottom) normalized returns, 
      before (left panel) and after (right panel) crash, given in 
      Eq.\,(\ref{eq:moment}) for $\Delta t=1,2,3,4,6,9,13$, and 19 days
      compared with those obtained from the Gaussian distribution
      (solid thick lines).}
   \label{fig:momenti} 
\end{figure}

\subsection{Probability Density Function}
Following previous studies \cite{mantegna1995,mantegna1999} we investigate
the probability density function [PDF or $P(Z)$] 
of the return to the origin $P(0)$ in order to 
investigate the statistical properties of both indices. 
The advantage of using such analysis is that we can
reduce the statistical inaccuracies, since the number of data 
included is relatively small, whereas the probability 
is largest at $Z=0$. 

 Starting with the characteristic function \cite{mantegna1999}
\begin{eqnarray}
  \label{eq:cfunction}
  \varphi(\Delta t) ~=~ e^{-\gamma \Delta t|q|^\alpha} ~,
\end{eqnarray}
the L\'evy stable distribution is given by
\begin{eqnarray}
  \label{eq:levy}
  P (Z) ~=~ \frac{1}{\pi}\, \int_{0}^{\infty}\, 
  e^{-\gamma \Delta t|q|^\alpha} \cos (qZ)\, dq ~.
\end{eqnarray}
From Eq.\,(\ref{eq:levy}) the probability of return to the origin $P(0)$ reads 
\begin{eqnarray}
  \label{eq:pz0}
  P(0) ~=~ \frac{\Gamma (1/\alpha)}{\pi\alpha(\gamma \Delta t)^{1/\alpha}} ~,
\end{eqnarray}
where $\Gamma$ indicates the Gamma function. 

The log-log plots of $P(0)$ as a function of the sampling 
time $\Delta t$ for both IHSG and KLSE indices are shown in 
the first and third column panels of Fig.\,\ref{fig:distxyf0}. 
The slopes of linear regressions to these plots equal the negative 
inverse of the L\'evy stable distribution indices $\alpha$. Using 
this index we calculate the parameter $\gamma$ by means of 
Eq.\,(\ref{eq:pz0}) and plot the ``theoretical'' PDF as a function 
of normalized returns $Z/\sigma$, using Eq.\,(\ref{eq:levy}), 
where $\sigma$ is the standard deviation of the distribution, 
and compare it with the empirical PDF obtained from data in 
Fig.\,\ref{fig:distxyf0}. 
Figure\,\ref{fig:distxyf0} reveals the fact that in the case of
IHSG the distribution turns from L\'evy to Gaussian 
after the crash. Surprisingly, the KLSE stock index shows 
a contrary result, the distribution alters from a Gaussian to 
a L\'evy one after the crash.

\subsection{Distribution Moment}
It has been argued that the use of the return probability
to the origin ${\rm PDF}(0)$ to estimate the L\'evy stable distribution 
index $\alpha$ is statistically not optimal, due to discreteness of the
distribution \cite{gopi1999}. Instead of exploiting such method, 
a different strategy has been proposed, i.e. by calculating  $\alpha$ by means of 
the slope of the cumulative distribution tails in a log-log plot.
To further test their results on the scaling behavior, 
Refs.\,\cite{gopi1999,plerou1999} analyzed the moments of the distribution
of normalized returns
\begin{eqnarray}
  \label{eq:moment}
  \mu (k) ~=~ \langle\, |g(t)|^k \rangle ~.
\end{eqnarray}
where the normalized returns $g(t)$ is defined by 
\begin{eqnarray}
  \label{eq:return_normalized}
  g(t) ~=~ \frac{Z(t)-\langle Z(t) \rangle}{\sqrt{\langle Z^2(t) \rangle
    - \langle Z(t) \rangle^2}} ~,
\end{eqnarray}
with $\langle Z(t) \rangle$ the time average of $Z(t)$
over the entire of time series. In the case of the S\&P 500 index
the result is found to be consistent with the analysis of the tails of
cumulative distributions. 
References\,\cite{gopi1999,plerou1999} pointed out that the change in 
the moments behavior originates from the gradual disappearance of
the L\'evy slope in the distribution tails. In our case it is also 
important to cross-check the result shown in Fig.\,\ref{fig:distxyf0}, 
since the number of data used in this analysis
is significantly smaller than that of previous analysis on the S\&P 500 
index, e.g. Ref. \cite{gopi1999}, which used approximately $5\times 10^6$ data points.
 

 It has been also shown in Ref.\,\cite{gopi1999} that Eq.\,(\ref{eq:moment}) 
will diverge for $k\ge 3$. In this study we also constrain $k$ 
within $0\le k\le 3$. The results for both indices compared with 
the moment obtained from a Gaussian distribution  have been reported 
in Ref. \cite{mart_econo_report}. It is found that the IHSG moment retains 
its scaling up to $\Delta t=9$ days, only after $\Delta t=13$ the moment starts to 
deviate toward the Gaussian distribution. In the KLSE case the moment quickly
converges to the Gaussian distribution as soon as $\Delta t$ increases from 1 day 
and does not show any scaling behavior as in the former case.

 In Fig.\,\ref{fig:momenti} we 
display the moments of both stock indices in the case of before
and after the financial crash. A consistent result, compared to the
analysis of Fig.\,\ref{fig:distxyf0}, 
is obtained from Fig.\,\ref{fig:momenti}, 
i.e., in the case of IHSG the distribution becomes closer
to Gaussian after crash, whereas the KLSE moments move away
from the Gaussian distribution after the crash.

 Another important finding obtained from Fig.\,\ref{fig:momenti} is that the
scaling behavior up to $\Delta t=9$ shown by the IHSG case in the 
previous section originates from the ``before crash'' period. After the crash,
the IHSG moments quickly converge to a Gaussian distribution. In fact,
this phenomenon has already been seen in the left panels of 
Fig.\,\ref{fig:distxyf0}, where the empirical 
probability of return to the origin is more scattered after crash.

\section*{Acknowledgments}
The author thanks Rachmat W. Adi and Suharyo Sumowidagdo 
for critical reading of the manuscript, as well as to 
Anto Sulaksono for useful discussions on the neutron star 
and for his contributions in part of the works
explained in this paper. 
This work has been partly supported by the
Hibah Pascasarjana grant and the Faculty of
Mathematics and Sciences, University of Indonesia.

\end{document}